\documentclass[aps,reprint,groupedaddress,longbibliography]{revtex4-1}

\usepackage{amsmath}
\usepackage{amsthm}
\usepackage{amsfonts}
\usepackage{braket}
\usepackage{graphicx}
\usepackage{wrapfig}
\usepackage{xcolor}

\DeclareMathOperator{\ops}{\textrm{ops}}

\DeclareMathOperator{\dd}{\text{d}\!}

\newcommand{\cfig}[1]{Fig.~\ref{#1}}
\newcommand{\ceqn}[1]{Eq.~(\ref{#1})}
\newcommand{\capp}[1]{Appendix~\ref{#1}}
\newcommand{\cref}[1]{Ref.~\cite{#1}}

\newcommand{\ctab}[1]{Table~\ref{#1}}
\newcommand{\cact}[1]{\hyperref[#1]{#1}}
\newcommand{\ignore}[1]{}

\begin{document}


\title{Automatic Differentiable Numerical Renormalization Group}

\author{Jonas B. Rigo}
\email[]{jonas.rigo@ucdconnect.ie}
\affiliation{School of Physics, University College Dublin, Belfield, Dublin 4, Ireland}
\affiliation{Centre for Quantum Engineering, Science, and Technology, University College Dublin, Belfield, Dublin 4, Ireland}

\author{Andrew K. Mitchell}
\email[]{andrew.mitchell@ucd.ie}
\affiliation{School of Physics, University College Dublin, Belfield, Dublin 4, Ireland}
\affiliation{Centre for Quantum Engineering, Science, and Technology, University College Dublin, Belfield, Dublin 4, Ireland}


\begin{abstract}
	Machine learning techniques have recently gained prominence in physics, yielding a host of new results
	and insights. One key concept is that of backpropagation, which computes the exact gradient of any output
	of a program with respect to any input. This is achieved efficiently within the differentiable programming
	paradigm, which utilizes automatic differentiation (AD) of each step of a computer program and the chain
	rule. A classic application is in training neural networks. Here, we apply this methodology instead to the
	numerical renormalization group (NRG), a powerful technique in computational quantum many-body physics.
	We demonstrate how derivatives of NRG outputs with respect to Hamiltonian parameters can be accurately and
	efficiently obtained. Physical properties can be calculated using this differentiable NRG scheme—for example,
	thermodynamic observables from derivatives of the free energy. Susceptibilities can be computed by adding
	source terms to the Hamiltonian, but still evaluated with AD at precisely zero field. As an outlook, we briefly
	discuss the derivatives of dynamical quantities and a possible route to the vertex.
\end{abstract}
\maketitle


\section{Introduction}
The Numerical Renormalization Group (NRG) \cite{wilson1975renormalization,bulla2008numerical}
is a standard tool in computational physics for solving a certain class of quantum many-body problem, known as \textit{quantum impurity models}. These comprise a few interacting quantum degrees of freedom, coupled to one or more non-interacting baths. The most famous example is the Kondo model, which involves a single spin-$\tfrac{1}{2}$ `impurity' coupled to a single fermionic bath \cite{hewson1997kondo}. The low-energy physics is non-perturbative and controlled by a single emergent energy scale (the Kondo temperature, $T_{\rm K}$), below which the impurity is dynamically screened by a spatially-extended entanglement `Kondo cloud' of surrounding conduction electrons \cite{mitchell2011real}. Generalized quantum impurity models describe the scattering from magnetic impurities in metals \cite{hewson1997kondo,costi2009kondo}, semiconductor quantum dot devices \cite{goldhaber1998kondo,keller2014emergent,iftikhar2015two,*mitchell2016universality,*iftikhar2018tunable,pouse2021exotic}, single molecule transistors \cite{liang2002kondo,mitchell2017kondo}, and are the local effective models within dynamical mean field theory (DMFT) for correlated materials \cite{kotliar2006electronic,georges1996dynamical,stadler2015dynamical}.

NRG is a tensor network method \cite{weichselbaum2009variational} that exploits the renormalization group structure of such quantum impurity models \cite{wilson1975renormalization}. At its core, the method involves the iterative numerical diagonalization of a set of renormalized effective Hamiltonians. There is no statistical element, as with quantum Monte Carlo impurity solvers \cite{gull2011continuous}.

In this paper, we apply the `differentiable programming' methodology to NRG, and demonstrate some advantages of its use for practical physics applications. We show that the core eigensolver routine in NRG is well-suited to Automatic Differentiation (AD), an emerging programming paradigm that has its origin in Deep Learning (DL) \cite{lecun2015deep}. In DL, hugely complicated neural networks (NN) feed their output into a loss function, which must be optimized with respect to the NN parameters (weights). This can be accomplished efficiently via gradient descent using AD programming: the gradient of the loss-function with respect to the NN weights can be computed at about the cost of the evaluation of the NN \cite{bartholomew2000automatic}. The remarkable aspect of this technique is that the computed gradient is not approximated, but is numerically exact. Without AD, derivatives can of course be approximated using finite difference (FD) derivatives. However, without care, FD derivatives can be inaccurate; when treated rigorously to guarantee convergence, FD can become computationally expensive. 
The efficiency and accuracy of AD owes to the fact that if each step of a computer program can be differentiated, then outputs can be differentiated with respect to inputs via the chain rule. Evaluating derivatives is trivial and cheap since the program itself is differentiated.

AD has been made more accessible through the introduction of AD libraries such as \textit{jax} \cite{jax2018github}, \textit{tensorflow} \cite{tensorflow2015} or \textit{PyTorch} \cite{NEURIPS2019_9015}. The advantage for computational physics is that when formulated using an AD library, exact derivatives can be obtained `for free' \cite{tamayo2018automatic,zhang2019automatic,xie2020automatic,mcgreivy2021optimized,ponsioen2021automatic,dick2021using,coopmans2021protocol}. Recent applications in physics include optimal control in quantum systems  \cite{vargas2021fully}, mitigating the sign problem in Monte Carlo \cite{wan2020mitigating}, and optimizing tensor networks \cite{liao2019differentiable}. In addition to its use for optimization problems, AD has particular appeal in physics because physical quantities are related to derivatives of generating functionals \cite{tu2021generating}.

Here we formulate a differentiable NRG ($\partial$NRG) scheme. The key step is an  explicit expression for the derivative of NRG eigenvalues and eigenvectors: the exact gradient of the entire eigensolver routine at any given NRG step is determined symbolically. This eliminates the need for full backpropagation, and when combined with AD for the other program elements, provides highly efficient access to derivatives of program outputs with respect to model parameters. For example, thermodynamic observables may be obtained from derivatives of the free energy; and the physical response of a system to perturbations can be simply and cheaply determined.


\section{Forward- and backward-mode AD}
 The AD approach is similar to symbolic differentiation, with standard derivative rules applied algorithmically \cite{linnainmaa1976taylor,lecun2015deep}. Regarding a program as a function $f$ that maps a given input $x$ to an output $y$, we may write $f(x) = y$. This is typically achieved in practice by concatenating several primitive functions $f^{(i)}$, viz:
\begin{equation}
\label{eq:simplegraph}
f = f^{(n)}\circ f^{(n-1)}\circ ... \circ f^{(1)} \; ,
\end{equation}
where ``$\circ$'' denotes the composition of two functions. We denote the number primitive functions comprising $f$ as $\ops(f)$. The flow of data between primitive functions can be expressed as a compute graph \cite{tinhofer2012computational}; for simplicity, we consider here only programs whose graph has a chain topology as in \ceqn{eq:simplegraph}.  To compute the derivative of $f$ with respect to $x$ one can apply the chain rule,
\begin{align}
\label{eq:chainrule}
\frac{\partial f}{\partial x} &= \frac{\partial f^{(n)}}{\partial f^{(n-1)}}\frac{\partial f^{(n-1)}}{\partial f^{(n-2)}}...\frac{\partial f^{(1)}}{\partial x} \;. 
\end{align}
\ceqn{eq:chainrule} can be evaluated algorithmically in either the \textit{forward mode}, or the \textit{backward mode}. In the forward mode, we apply the recursion 
\begin{equation}
\label{eq:forward}
\frac{\partial f^{(i)}}{\partial x} = \frac{\partial f^{(i)}}{\partial f^{(i-1)}}\delta_{i-1}\equiv \delta_{i},~\textrm{with}~\delta_{0} = 1,~f^{(0)} = x
\end{equation}
starting with $i=1$ and increasing up to $i=n$, with $f^{(n)}\equiv f$ such that $\delta_n=\partial f/\partial x$ is the desired derivative. In backward mode (also known as \textit{backpropagation}) we use the recursion 
\begin{equation}
\label{eq:backward}
\frac{\partial f}{\partial f^{(i)}} = \frac{\partial f^{(i+1)}}{\partial f^{(i)}}\bar{\delta}_{i+1}\equiv \bar{\delta}_{i},~\textrm{with}~\bar{\delta}_{n} = 1,~f^{(0)} = x 
\end{equation}
starting from $i=n-1$ and decreasing down to $i=0$, with $\bar{\delta}_0=\partial f/\partial x$ the desired derivative.

 These forward and backward mode AD recursions are symbolically equivalent methods for calculating derivatives. In the practical implementation of AD the intermediate derivatives $\delta_i$ or $\bar{\delta}_i$ are evaluated numerically and stored in memory. This provides a way to compute numerically exact derivatives, with the true analytical result recovered when the value of $\delta_i$ is not truncated \cite{griewank2008evaluating}.

By contrast, the FD approximation reads,
\begin{equation}
\label{eq:FD1}
\frac{\partial f}{\partial x} \approx \frac{f(x+h) -f(x)}{h} \equiv D_h[f](x) \; ,
\end{equation}
with $h$ being the FD value. 
The exact limit $h \rightarrow 0^{\pm}$ cannot be taken numerically, and so $\partial_x f$ must be approximated for finite $h$ and the convergence  $D_{h}[f](x) \rightarrow \partial_x f$ checked explicitly. This may require many function evaluations (runs of the whole program) since for very small $h$, $f(x)$ and $f(x+h)$ may be numerically indistinguishable  (truncation error). Furthermore, convergence should be checked for both the forwards ($h>0$) and backwards ($h<0$) difference quotient. The situation gets worse for higher-order derivatives, and numerical convergence of FD approximations can be costly and/or fraught \cite{grossmann2007numerical}.

Above we considered scalar functions of the form $f: \mathbb{R} \rightarrow \mathbb{R}$. However, both AD and FD approaches can be straightforwardly generalized to compute the Jacobian of vector-valued functions defined over a vector space $f: \mathbb{R}^n \rightarrow \mathbb{R}^m$. The FD and forward/backward mode AD scale differently with respect to the input dimension $n$ and output dimension $m$. The FD exhibits the most expensive scaling of computational cost with $\sim 2 n m$ function evaluations required to compute the Jacobian of $f$ in the single-shot case. By contrast, AD methods do not require the function to be evaluated in order to compute the derivative, as with the symbolic approach (although since many derivatives are related to their anti-derivative, function evaluations can be reused to reduce computational cost at the expense of increased memory cost \cite{baydin2018automatic}). Instead, the computational cost of AD methods is controlled by $\ops(f)$. Specifically, the forward mode scales as $\sim c_F n \ops(f)$, while the backward mode scales as $\sim c_B m \ops(f)$, with $c_F, c_B < 6$ (although the forward mode typically outperforms the backward mode unless $m \ll n$ \cite{baydin2018automatic}). The computational cost of the FD approach is therefore typically much higher than either of the AD modes; in this work we use forward mode AD unless stated otherwise. A comparison of the performance and precision of the different methods, as applied to NRG, is presented in \capp{app:benchmark}.

Through AD one can obtain the numerically exact derivative of any program with a single run, as long as all operations comprising the program are themselves differentiable. Thus, one can differentiate a given physics solver program simply by utilizing known derivatives of its constituent operations. 

On the other hand, in certain cases it may be possible to find the symbolic derivative of the entire solver with respect to certain input parameters. In the following we show how the latter can be achieved for NRG through analysis of the backpropagation chain.


\section{NRG}
Wilson's NRG is a numerical solver for quantum impurity models, of the type $\hat{H}=\hat{H}_{\rm imp} + \hat{H}_{\rm bath} + \hat{H}_{\rm hyb}$, where $\hat{H}_{\rm imp}$ is the `impurity' Hamiltonian describing a few local, interacting quantum degrees of freedom, while $\hat{H}_{\rm bath} = \sum_{\sigma,\mathbf{k}}\epsilon_{\mathbf{k}}^{\phantom{\dagger}}\hat{c}^\dagger_{\sigma\mathbf{k}}\hat{c}_{\sigma\mathbf{k}}^{\phantom{\dagger}}$ describes a non-interacting fermionic bath. The coupling between them is given by $\hat{H}_{\rm hyb} =\sum_{\sigma,\mathbf{k}} V_\mathbf{k}^{\phantom{\dagger}}(\hat{d}^\dagger_{\sigma}\hat{c}_{\sigma\mathbf{k}}^{\phantom{\dagger}} + \hat{c}^\dagger_{\sigma\mathbf{k}}\hat{d}_{\sigma}^{\phantom{\dagger}})$, where we have assumed for simplicity here that a single impurity orbital $\hat{d}_{\sigma}$ couples to the bath.  

Due to the impurity interactions, such a model is a genuine quantum many body problem, and in general has no exact solution \cite{hewson1997kondo}. NRG treats instead a discretized version of the model, which becomes computationally tractable through a process of iterative diagonalization and truncation. This constitutes an RG procedure, in which useful physical information can be extracted at each step, as progressively lower energy scales are probed. Full details can be found in Refs.~\cite{wilson1975renormalization,bulla2008numerical}; here we introduce only the elements necessary to formulate $\partial$NRG.

The first step is the logarithmic discretization of the bath and mapping to a Wilson chain. The continuous local density of states of the uncoupled bath at the impurity position is divided up into intervals of decreasing width, defined by the points $x_n^\pm = \pm D \Lambda^{-n}$, where $D$ is the half-bandwidth, $\Lambda>1$ the discretization parameter, and $n=0,1,2,...$. The spectrum is discretized by replacing the continuous density in each interval by a single pole of the same weight and at the average position. A semi-infinite tight-binding chain (the Wilson chain) is then defined such that the spectral function at the end is the same as the discretized bath spectrum (this is achieved in practice by the Lanczos algorithm). The discretized bath Hamiltonian reads,
\begin{equation}
\label{eq:wc}
\hat{H}_{\rm bath}~\mapsto~\hat{H}_{\rm bath}^{\rm disc} =\sum_{\sigma}\sum^{\infty}_{n=0}t_n(\hat{c}^\dagger_{\sigma n+1}\hat{c}_{\sigma n} + \hat{c}^\dagger_{\sigma n}\hat{c}_{\sigma n+1}) \; ,
\end{equation}
where we have assumed particle-hole symmetry here for simplicity. The original model is recovered as $\Lambda \rightarrow 1$. The specific form of the Wilson chain hopping parameters $t_n$ depends on details of the dispersion relation $\epsilon_{\mathbf{k}}$. However, for a metallic system, $t_n \sim D \Lambda^{-n/2}$ for large $n$. 

For a Wilson chain of total length $N$, adding an extra site $N+1$ is therefore always a small perturbation. The NRG exploits this energy scale separation down the chain through the iterative diagonalization scheme. Starting from the impurity, the chain is built up by successively adding new Wilson chain sites. At each step, the system is diagonalized, and high-energy states discarded. One defines a sequence of rescaled  Hamiltonians, comprising the impurity and the first $N$ sites of the Wilson chain, 
\begin{align}
\label{eq:HN}
\hat{H}_N = \Lambda^{(N-1)/2}\left[\hat{H}_{\text{imp}} + V \sum_{\sigma}(\hat{d}^\dagger_{\sigma }\hat{c}_{\sigma 0} + \hat{c}^\dagger_{\sigma 0}\hat{d}_{\sigma }) 
\right. \nonumber \\
\left. + \sum_{\sigma}\sum^{N-1}_{n=0}t_n(\hat{c}^\dagger_{\sigma n+1}\hat{c}_{\sigma n} + \hat{c}^\dagger_{\sigma n}\hat{c}_{\sigma n+1})\right] \; ,
\end{align}
with $V$ the effective impurity-bath coupling. The full (discretized) Hamiltonian is recovered in the limit $\hat{H} = \lim_{N\rightarrow\infty}\Lambda^{-(N-1)/2}\hat{H}_N$. The sequence of Hamiltonians $\hat{H}_N$ satisfy the recursion relation, 
\begin{equation}
\label{eq:HN+1}
\hat{H}_{N+1} = \sqrt{\Lambda}\hat{H}_N + \Lambda^{N/2}\sum_\sigma t_N(\hat{c}^\dagger_{\sigma N+1}\hat{c}_{\sigma N} + \hat{c}^\dagger_{\sigma N}\hat{c}_{\sigma N+1}) \; .
\end{equation} 

At each step $N$, the Hamiltonian $\hat{H}_N$ is diagonalized to find the eigenvectors (states)  $\{|i\rangle_N\}$ and eigenvalues (energies) $\{E_{N;i}\}$ that satisfy the Schr\"odinger equation,
\begin{equation}
\hat{H}_N \ket{i}_N = E_{N;i}\ket{i}_N \; .
\end{equation}
We denote excitation energies relative to the ground state as $\Delta E_{N;i} = E_{N;i}-E_{N;0}$, and $\hat{H}_{\rm imp}\equiv \hat{H}_{-1}$. 
To construct $\hat{H}_{N+1}$ we add another Wilson chain site. The Fock space of $\hat{H}_{N+1}$ is spanned by basis states,
\begin{equation}
\label{eq:prodbasis}
\ket{i;s}_{N+1} = \ket{i}_N\otimes \ket{s}\; ,
\end{equation}
comprising the tensor product of eigenstates $|i\rangle_N$ of $\hat{H}_N$, and the added chain site $N+1$ denoted $|s\rangle$. Matrix elements of $\hat{H}_{N+1}$ in this basis read,
\begin{equation}
\label{eq:grownbasis}
\hat{H}_{N+1}(is;i's') = \;_{N+1}\!\bra{i;s} \hat{H}_{N+1} \ket{i';s'}_{N+1} \;.
\end{equation}
Finally, $\hat{H}_{N+1}$ is diagonalized to find the new eigenbasis at step $N+1$, viz:
\begin{equation}
\label{eq:basistrafo}
\ket{j}_{N+1} = \sum_{is}U_{N+1}(j,is)\ket{i;s}_{N+1} \; .
\end{equation}
 \ceqn{eq:grownbasis} can be simplified using the energies $E_{N;i}$ of $\hat{H}_{N}$ and the tensor $\eta_{ss'\sigma} = \bra{s'} \hat{c}_{\sigma N+1} \ket{s}$ (which is independent of $N$), viz
	\begin{align}
	\label{eq:newH}
	\hat{H}_{N+1}(is;i's') = &\sqrt{\Lambda}\delta_{ss'}\delta_{ii'}\times E_{N;i} \\ &+(-1)^s\Lambda^{N/2}t_N\sum_{\sigma} \eta_{ss'\sigma} \eta_{N;ii'\sigma} + {\rm H.c.} \; ,\nonumber
	\end{align}
where $\eta_{N;jj'\sigma} =  \sum_{\bar{i}\bar{s},\bar{i}'\bar{s}'} U^\dagger_{N}(j,\bar{i}\bar{s}) ~U_{N}(j',\bar{i}'\bar{s}') \times \delta_{\bar{i}\bar{i}'}\eta_{\bar{s}'\bar{s}\sigma}$ and $s=\{0,-1,+1,2\}$ for $|s\rangle=\{|0\rangle,|\downarrow\rangle,|\uparrow\rangle,|\uparrow\downarrow\rangle\}$.

When following these steps, the dimension of the Fock space grows exponentially with the length of the chain. This is avoided in NRG by truncating the Fock space at each step, discarding high-energy states. In practice, one retains $M_K$ of the lowest energy eigenstates of $\hat{H}_N$ at each step. The \textit{NRG approximation} \cite{wilson1975renormalization,weichselbaum2007sum} is that the states and energies of $\hat{H}_N$ approximate those of the full $\hat{H}$. This approximation is justified by the exponential decay of the Wilson chain coefficients $t_n$, which means that high-energy states discarded at a given iteration do not become low-lying states at a later iteration. Convergence of the NRG calculation can be checked \emph{post hoc} by increasing $M_K$. Hereafter it is to be understood that the condition $i\le M_K$ applies to \ceqn{eq:prodbasis} (with the eigenvalues sorted lowest to highest). The dimensionality of the Fock space is therefore constant at each step, and the NRG calculation scales \emph{linearly} with chain length $N$. With increasing $N$ the physics on successively lower energy scales is uncovered.

The entire process of going from one iteration to the next can be summarized as an RG transformation, 
\begin{equation}
\label{eq:RG}
\hat{H}_{N+1} = R[\hat{H}_N] \; .
\end{equation}
 The full (discretized, renormalized) Hamiltonian $\hat{H}_N$ can therefore be constructed iteratively starting from the impurity Hamiltonian $\hat{H}_{\rm imp}$. Regarding the latter as a function of $n$ physical model parameters $\{\theta \}$, the entire NRG algorithm can be viewed as a function,
\begin{align}
\label{eq:NRGdim}
	&f(\lbrace \theta \rbrace) \equiv R^{\circ N+1}[\hat{H}_{\rm imp}(\lbrace \theta \rbrace)]  \nonumber \\
	&f:\mathbb{R}^n \rightarrow \mathbb{C}^{M_K\times M_K}\; ,
\end{align}
where we have used the shorthand notation for an $N\!+\!1$-fold concatenation of functions $R^{\circ N+1} = R \circ R \circ ... \circ R$. All physical quantities of interest for the original quantum impurity model are obtained from the eigenvalues and eigenvectors of the set of $\{\hat{H}_N\}$ \cite{wilson1975renormalization,bulla2008numerical,weichselbaum2007sum}. 

Note that an operator $\hat{O}$, acting only on impurity degrees of freedom, can be expressed in the in the eigenbasis of $\hat{H}_{N}$. We denote this matrix as $\hat{O}_N(i;i')=\;_{N}\!\bra{i} \hat{O}\ket{i'}_{N}$. From \ceqn{eq:basistrafo} it follows that,
\begin{align}
\label{eq:Ptrafo}
&\hat{O}_{N+1}(j;j') \nonumber\\ &=\sum_{is,i's'}\!\!{\vphantom{\sum}}'
U^\dagger_{N+1}(j,is) ~U_{N+1}(j',i's') \times \hat{O}_{N+1}(is;i's') \nonumber\\
& =\sum_{is,i's'}\!{\vphantom{\sum}}' U^\dagger_{N+1}(j,is) ~U_{N+1}(j',i's') \times \hat{O}_{N}(i;i')\delta_{s,s'}
\end{align}
where the primed sum implies that $i,i' \le M_K$, and the second line follows from the fact that $\hat{O}$ does not act on states $|s\rangle$ of site $N+1$. \ceqn{eq:Ptrafo} can be used iteratively to transform $\hat{O}_{-1}(i;i')$, evaluated explicitly in the impurity basis, into the eigenbasis of any later iteration. 

Symbolically, we denote this process,
\begin{equation}
\label{eq:Pop}
\hat{O}_{N+1} = P[\hat{O}_N] \; ,
\end{equation}
which we refer to as \textit{propagating forward} the operator.


\section{Derivative of eigenvalues and eigenvectors in NRG}
As outlined in the previous section, NRG outputs the eigenvalues and eigenvectors for the set of $\hat{H}_N$. Since all physical quantities for the quantum impurity model are obtained from these, calculating their derivatives using \ceqn{eq:chainrule} requires the derivatives of the eigenvalues and eigenvectors of $\hat{H}_N$.  From \ceqn{eq:NRGdim}, we regard NRG as a function defined over an $n$ dimensional domain, with an $M_K\times M_K$ dimensional image. Since $n$ denotes the number of model parameters in $\hat{H}_{\rm imp}$, and $M_K \times M_K$ is the Fock space dimension of $\hat{H}_{N}$, we have $M_K \times M_K \gg n$ and therefore forward mode AD is significantly faster \cite{griewank2008evaluating}. In the following we consider only forward mode AD.

For a Hermitian matrix $\hat{A}$ (such as the Hamiltonian $\hat{H}_N$) with non-degenerate eigenvectors $|i\rangle$ and distinct eigenvalues $\lambda_i$ satisfying $\hat{A}|i\rangle=\lambda_i|i\rangle$, we may express the differentials through  \cite{seeger2017auto,giles2008extended},
\begin{subequations}
	\label{eq:Aderivs}
	\begin{align}
	\dd \lambda_i &= \bra{i} \dd \hat{A} \ket{i} \; , \label{eq:deiv} \\
	\dd\ket{i} &= \sum_{i\neq j}\frac{\bra{j} \dd \hat{A} \ket{i}}{\lambda_i - \lambda_j}\ket{j} \; .\label{eq:deis}
	\end{align}
\end{subequations}
These expressions are of course well-known in the context of non-degenerate perturbation theory  \cite{feynman1939forces}.

However, the assumption implicit in \ceqn{eq:Aderivs} that $\hat{A}$ must be free of eigenvalue degeneracies is rather restrictive.
Methods treating the general degenerate case are much more computationally involved and may suffer from numerical instabilities  \cite{dailey1989eigenvector,giles2008extended}. Of course in physics, where $\hat{A}$ is the Hamiltonian of some system, energy eigenvalue degeneracies are common. This may be because of underlying non-Abelian symmetries (for example SU(2) spin symmetry), which endow the spectrum of the Hamiltonian with a multiplet structure. In this case, the Hamiltonian becomes block diagonal in the associated conserved quantum numbers, and the diagonalization \ceqn{eq:basistrafo} can be done separately in each block. Alternatively, NRG can be formulated directly in multiplet space by using the Wigner-Eckart theorem \cite{weichselbaum2012non}. Either approach removes the problem of such degeneracies in \ceqn{eq:Aderivs}. However, this may not fully solve the problem, since there could be accidental degeneracies or emergent symmetries \cite{wilson1975renormalization,hewson1997kondo,keller2014emergent} that lead to additional energy eigenvalue degeneracies in physical systems. 

To overcome this, we make a simple approximation. To the diagonal entries of $\hat{H}_N$ we add noise, with random variables drawn from a  normal distribution of width $\sigma$. The noise lifts the eigenvalue degeneracy, meaning that derivatives can be obtained using \ceqn{eq:Aderivs}. Note that the smallest terms in the rescaled Hamiltonian $\hat{H}_N$ are $\mathcal{O}(1)$  \cite{wilson1975renormalization,bulla2008numerical}, so $\sigma \ll 1$ constitutes a small perturbation.
Care should be taken to add the noise in such a way to respect bare symmetries.

 With a straight application of AD via \ceqn{eq:forward}, the derivatives of eigenvalues and eigenvectors for $\hat{H}_N$ require the evaluation of \ceqn{eq:Aderivs} at every NRG step (and hence noise must be added at every NRG step). For small enough noise width $\sigma$, physical properties at a given iteration should be unaffected. Indeed, in \capp{app:benchmark} we show for the AIM that NRG remains highly accurate when  $\sigma \lesssim 10^{-6}$. At larger noise levels, errors may propagate through the iterative process (and will snowball if they introduce RG relevant perturbations), so care must be taken to avoid this with such an AD approach. 
On the other hand, forward mode AD for the AIM requires $\sigma \gtrsim 10^{-6}$ to stabilize the calculation of derivatives using \ceqn{eq:Aderivs}. Smaller noise levels introduce more severe numerical instabilities because of the recursive nature of the derivative calculation in AD via \ceqn{eq:forward}  -- derivatives at one iteration depend on those of previous iterations. This stability-accuracy tradeoff in AD is analyzed in \capp{app:benchmark}.

 Below, we derive an alternative formulation used in $\partial$NRG, which utilizes the symbolic derivative of the entire eigensolver for $\hat{H}_N$. This has the advantage that \ceqn{eq:Aderivs} need only be applied once, at the NRG step for which derivatives are required. Not only is this far less computationally costly, but it also  avoids accumulating errors introduced by noise, which can then be added at a much lower level.


\subsection{Formulation in $\partial$NRG}
In the context of NRG, one can formulate a differential recursion relation based on the RG transformation \ceqn{eq:RG},
\begin{equation}
\label{eq:NRGfAD}
\dd\hat{H}_N = \dd \hat{H}_{N-1}\frac{\dd R(\hat{H}_{N-1})}{\dd \hat{H}_{N-1}} \; .
\end{equation}
This allows derivatives of $\hat{H}_N$ to be computed  in AD forward or backward mode. However, one can also reformulate the problem in a much more simple and efficient way. Taking derivatives with respect to model parameters $\theta$ of $\hat{H}_{\rm imp}$, we may write from \ceqn{eq:HN} the operator identity $\partial_{\theta} \hat{H}_N \simeq \Lambda^{(N-1)/2}~\partial_{\theta} \hat{H}_{\rm imp}$ (where $\simeq$ follows from the NRG approximation).  Since $\partial_{\theta} \hat{H}_{\rm imp}$ only involves impurity operators, it can be trivially evaluated in the impurity basis, $[\partial_{\theta} \hat{H}_{\rm imp}]_{-1}(k;k')={}_{-1}\langle k| \partial_{\theta} \hat{H}_{\rm imp} |k'\rangle_{-1}$. \ceqn{eq:Pop} can then be used to propagate this operator forward into the basis of $\hat{H}_N$. It therefore follows that,
\begin{equation}
\label{eq:dH}
[\partial_{\theta}\hat{H}_{N}]_N \simeq \Lambda^{(N-1)/2}~ P^{\circ N+1}([\partial_{\theta} \hat{H}_{\rm imp} ]_{-1}) \;.
\end{equation}
From \ceqn{eq:Aderivs} we then obtain,
\begin{subequations}\label{eq:mainresult}
	\begin{align}
	\label{eq:dd}
	\partial_{\theta} E_{N;i} &\simeq \Lambda^{(N-1)/2} ~ P^{\circ N+1}([\partial_{\theta} \hat{H}_{\rm imp}]_{-1}) (i;i)  \\
	\partial_{\theta}\ket{i}_N &\simeq \Lambda^{(N-1)/2} \sum_{i\neq j}\frac{ P^{\circ N+1}([\partial_{\theta} \hat{H}_{\rm imp}]_{-1}) (j;i)}{E_{N;i} - E_{N;j}}\ket{j}_N 
	\end{align}
\end{subequations}
\ceqn{eq:mainresult} is the main result of this work. It shows how the derivative of energies and states of the NRG Hamiltonians $\hat{H}_N$ can be obtained by simply forward-propagating the impurity operators $\partial_{\theta} \hat{H}_{\rm imp}$. Derivatives of physical quantities can then be related to those obtained via \ceqn{eq:mainresult}. The above can also be generalised to deal with derivatives $\partial_{\phi}$ of other Hamiltonian parameters $\phi$ (for example $V$ or $t_n$ in \ceqn{eq:HN})  and higher order derivatives (see \capp{app:secoder}).
	
 To calculate first-order derivatives of eigenvalues and eigenvectors in AD forward mode (assuming the simplest NRG implementation) one must perform seven additional operations for every NRG step. The $\partial$NRG approach involves only one additional operation per NRG step and does not require any propagation of derivatives related to the chain rule (see \capp{app:wegenert}). Hence $\partial$NRG, utilizing \ceqn{eq:mainresult}, is far more efficient than standard forward mode AD. Performance benchmarking is demonstrated in \ref{app:benchmark}.

 Derivatives with respect to multiple parameters can be obtained straightforwardly by propagating forward any $[\partial_{\theta} \hat{H}_{\rm imp} ]_{-1}$ of interest using \ceqn{eq:Ptrafo} in $\partial$NRG. Therefore, $\partial$NRG scales linearly with the number of derivatives $n$, but is independent of $\ops(f)$, unlike forward mode AD.

 In practice, the use of \ceqn{eq:mainresult} again requires lifting eigenvalue degeneracies by adding a small diagonal noise term. We found in numerical tests for the AIM that $\sigma\simeq10^{-11}$ is sufficient for stabilizing the numerics in $\partial$NRG, without noticeably impacting any measurable physical properties  -- see \capp{app:benchmark}. 

If derivatives at only step $N$ are required, such a noise term need only be introduced at that step (rather than adding noise at each step) and the RG flow is unaffected. This is typically the situation for ground-state properties or low-temperature thermodynamics. On the other hand, if information is required from every NRG step (for example in the calculation of dynamical quantities \cite{weichselbaum2007sum}) noise must be introduced at each iteration. Unlike with the straight AD implementation, $\partial$NRG allows us to mitigate the possibility of snowballing errors introduced by propagating the noise terms -- if this level of accuracy should be required (e.g.~in the vicinity of a quantum critical point). This is because the eigensystem derivatives at different $N$ via \ceqn{eq:mainresult} are independent in $\partial$NRG. Noise may be added to $\hat{H}_N$ for the purpose of evaluating \ceqn{eq:mainresult}, but the pristine $\hat{H}_N$ without noise can be used for the main NRG recursion. This gives the most accurate and reliable results, at the cost of an additional matrix diagonalization. See Appendices \ref{app:benchmark} and \ref{app:wegenert} for performance comparisons (and note that this more rigorous approach is still faster than FD and straight AD in many circumstances, and certainly more accurate).


\section{Application to Anderson Impurity Model}
We illustrate the use of \ceqn{eq:mainresult} by applying the $\partial$NRG scheme to the paradigmatic Anderson impurity model (AIM), for which
\begin{equation}
\label{eq:aim}
\hat{H}_{\text{imp}} = \epsilon \hat{n} +
U \hat{n}_{\uparrow}\hat{n}_{\downarrow} \;, 
\end{equation}
where $\hat{n} =\sum_{\sigma}\hat{n}_{\sigma}$ and $\hat{n}_{\sigma}=\hat{d}^{\dagger}_{\sigma}\hat{d}^{\phantom{\dagger}}_{\sigma}$. With $\hat{H}_{\rm imp}$ so defined, the exact (discretized) $\hat{H}_N$ is given by \ceqn{eq:HN}. In NRG, $\hat{H}_N$ is approximated through the RG procedure \ceqn{eq:RG}, $\hat{H}_N = R^{\circ N+1}(\hat{H}_{\text{imp}})$.

For a given $\hat{H}_N$, we obtain the partition function $\mathcal{Z}_N(\bar{\beta}) = \sum_{i} e^{-\bar{\beta} E_{N;i}}$ and free energy $F_N(\bar{\beta}) = -\frac{1}{\beta}\log[\mathcal{Z}_N(\bar{\beta})]$. 
In the original Wilsonian formulation \cite{wilson1975renormalization,bulla2008numerical}, the effective inverse temperature $\beta\equiv 1/k_{\rm B} T$  is related to the NRG iteration number (Wilson chain length) $N$ via $\beta = \Lambda^{(N-1)/2}\bar{\beta}$, with $\bar{\beta} =\mathcal{O}(1)$ in practice. With this definition, the NRG free energy at inverse temperature $\bar{\beta}$ is a good approximation to the true free energy of the original (undiscretized) model at inverse temperature $\beta$  \cite{wilson1975renormalization,bulla2008numerical}, $F(\beta) \simeq F_N(\bar{\beta})$. The corresponding differential follows as,
\begin{equation}
\label{eq:dF}
\dd F_N = \frac{\Lambda^{-(N-1)/2}}{\mathcal{Z}_N}\sum_{i} e^{-\bar{\beta} E_{N;i}} \dd E_{N;i} \;.
\end{equation}
For the AIM, we may use \ceqn{eq:mainresult} to obtain  derivatives of the free energy with respect to the impurity parameters $\epsilon$ and $U$. For example, 
\begin{equation}
\label{eq:FN_eps}
\partial_\epsilon F_N =  \frac{1}{\mathcal{Z}_N}\sum_i e^{-\bar{\beta} E_{N;i}} P^{\circ N+1}([\hat{n}]_{-1})(i;i)  \; .
\end{equation}
Since $\partial_\epsilon F_N(\bar{\beta}) = \langle \hat{n} \rangle_{H_N,\bar{\beta}}\simeq \langle \hat{n} \rangle_{H,\beta}$, \ceqn{eq:FN_eps} is precisely equivalent to the standard Wilsonian approach to calculating local thermodynamic expectation values in NRG \cite{bulla2008numerical}. The above illustration demonstrates that $\partial$NRG is analytically equivalent to the well-known result for such local thermodynamic quantities. 

 Another commonly computed quantity for such models is the \textit{local} impurity magnetic susceptibility at zero field,
	\begin{align}
	\chi(T) = \int^\beta_0 \dd\tau\langle \hat{S}_z(\tau)\hat{S}_z\rangle - \beta\langle \hat{S}_z\rangle^2 \nonumber 
	\end{align}
where $	\hat{S}_z = \tfrac{1}{2} (\hat{n}_{\uparrow} - \hat{n}_{\downarrow})$ is the impurity spin operator and $\tau$ is imaginary time. This can be alternatively obtained by adding a source term $B\hat{S}z$ to the Hamiltonian Eq.~\ref{eq:aim}, and then taking the second-order derivative of the free energy with respect to $B$, evaluated at $B=0$. That is, we can write $\chi(T)=\frac{\partial^2}{\partial B^2} F\big|_{B = 0} $. Since $\partial$NRG is able to deal with second (and higher) order derivatives (see \capp{app:secoder}) we calculate  $\chi_{H_N,\bar{\beta}}=\frac{\partial^2}{\partial B^2} F_N(\bar{\beta})\big|_{B = 0}$. This is obtained automatically in $\partial$NRG, but  from \ceqn{eq:2ndorder} it can also be expressed as,
	\begin{align}
	\label{eq:chi}
	&\chi_{H_N,\bar{\beta}} = -\frac{\bar{\beta}}{\mathcal{Z}^2_N}\left(\sum_i e^{-\bar{\beta} E_{i;N}} P^{\circ N+1}([\hat{S}_z]_{-1})(i;i)\right)^2 \nonumber \\
	&+ \frac{\bar{\beta}}{\mathcal{Z}_N}\sum_i e^{-\bar{\beta} E_{i;N}} |P^{\circ N+1}([\hat{S}_z]_{-1})(i;i) |^2 \nonumber  \\
	&- \frac{2}{\mathcal{Z}_N}\sum_i e^{-\bar{\beta} E_{i;N}}\sum_{i\neq j} \frac{| P^{\circ N+1}([\hat{S}_z]_{-1})(j;i)|^2}{E_{N;i} - E_{N;j}} \; .
	\end{align}
This form of $\chi_{H_N,\bar{\beta}}$ is equivalent to the Lehmann representation of $\chi(T)$ evaluated in NRG Hamiltonian $\hat{H}_N$ at effective inverse temperature $\bar{\beta} = \beta\Lambda^{-(N-1)/2}$ \cite{bulla2008numerical}. Since all degeneracies are lifted in $\partial$NRG, convergence factors typically used in the Lehmann representation of dynamical quantities are not needed here. Note that $\chi_{H_N,\bar{\beta}}$ can be obtained at exactly zero magnetic field, so no symmetries are broken and no limit is taken numerically. Since the derivatives of primitive program functions are symbolic, source terms added to $\hat{H}_{\rm imp}$ may be evaluated at zero coupling constant and still yield finite derivatives. However, we note that such symmetry-breaking source terms cannot be added to the Hamiltonian if non-Abelian quantum numbers are implemented from the outset. In the case of the magnetic susceptibility, we can therefore utilize conserved total $S_z$, but not conserved total $S$, when adding the source term $B\hat{S}_z$ (even though in the end we set $B=0$). Numerical results are presented in \capp{app:benchmark}, and reproduce precisely the results of standard NRG using \ceqn{eq:chi}.
	
 Similarly, $\partial$NRG may be used to obtain the charge susceptibility, which in the wide-band limit is given simply by $\chi_C(T)\simeq\partial^2 F_N(\bar{\beta})/\partial \epsilon^2 = \partial_{\epsilon} \langle \hat{n} \rangle_{H_N,\bar{\beta}}$ (and can be obtained by replacing $\hat{S_z}$ by $\hat{n}$ in \ceqn{eq:chi}). Indeed, Maxwell relations provide non-trivial connections between physical quantities obtained as derivatives. This was exploited recently in Ref.~\cite{han2021extracting} to extract the fractional entropy of multi-channel Kondo states in quantum dot experiments, comparing with NRG calculations of charge derivatives.

 The above examples for the AIM demonstrate that the $\partial$NRG algorithm is  equivalent to known expressions for certain derived quantities. However, the power of $\partial$NRG is that it does this automatically within a generalized framework, and works equally well for any derivative in any quantum impurity model.


\begin{figure}[t]
	\includegraphics[width=\columnwidth]{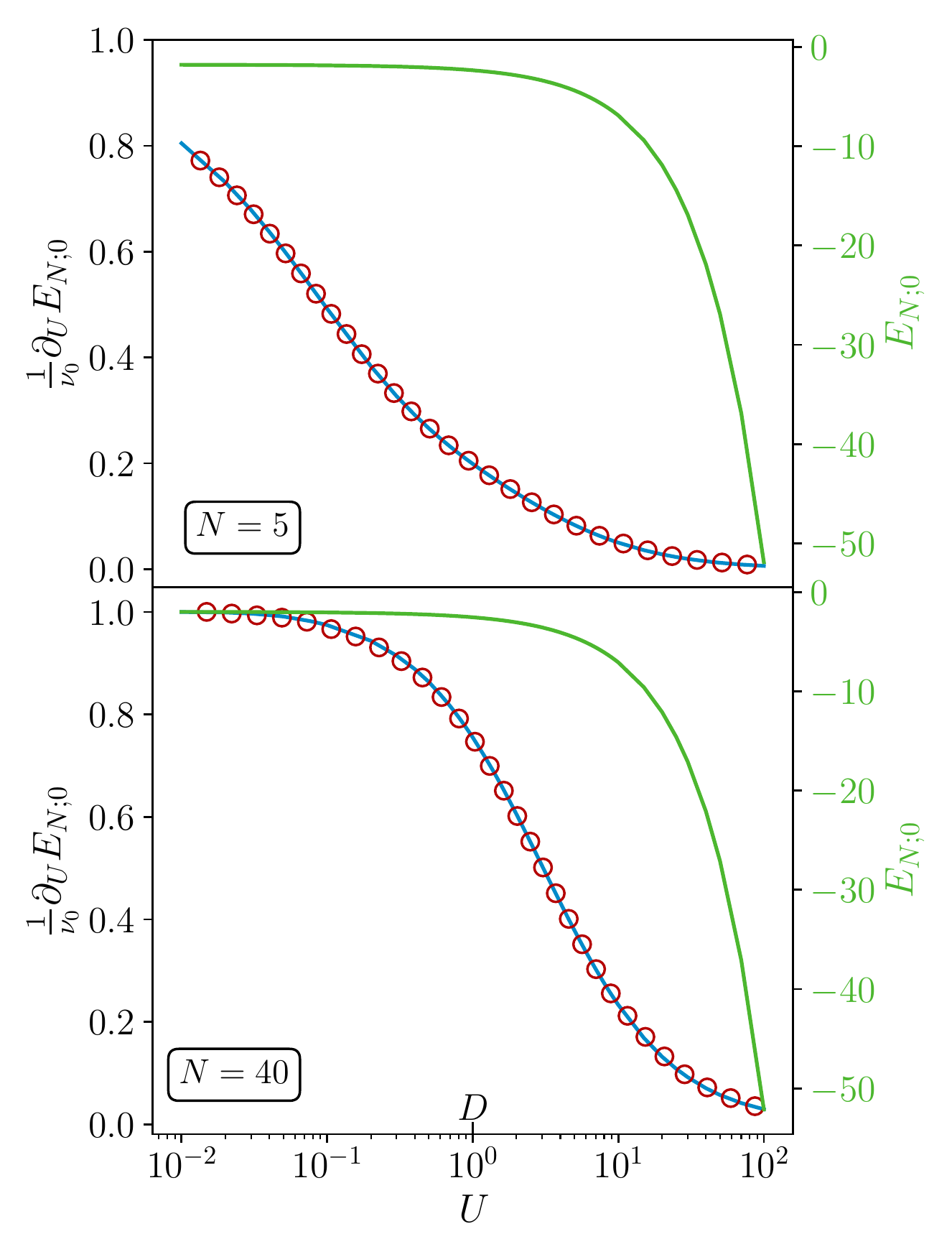}
	\caption{Ground state energy $E_{N;0}$ of the AIM (green line) and its normalized derivative $\partial_U E_{N;0}/\nu_0$ obtained by $\partial$NRG (blue line) and by FD (red circles) for iteration $N=5$ (top panel) and $N=40$ (bottom panel). Results plotted as a function of $U$, with $\epsilon=-U/2$ and constant $J\equiv 8V^2/U=0.3$. We define $\nu_0 = \partial_U E_{N;0}\vert_{U \rightarrow 0}$ as the derivative in the limit $U\to 0$.}
	\label{fig:dGS}
\end{figure}

\begin{figure}[t]
	\includegraphics[width=\columnwidth]{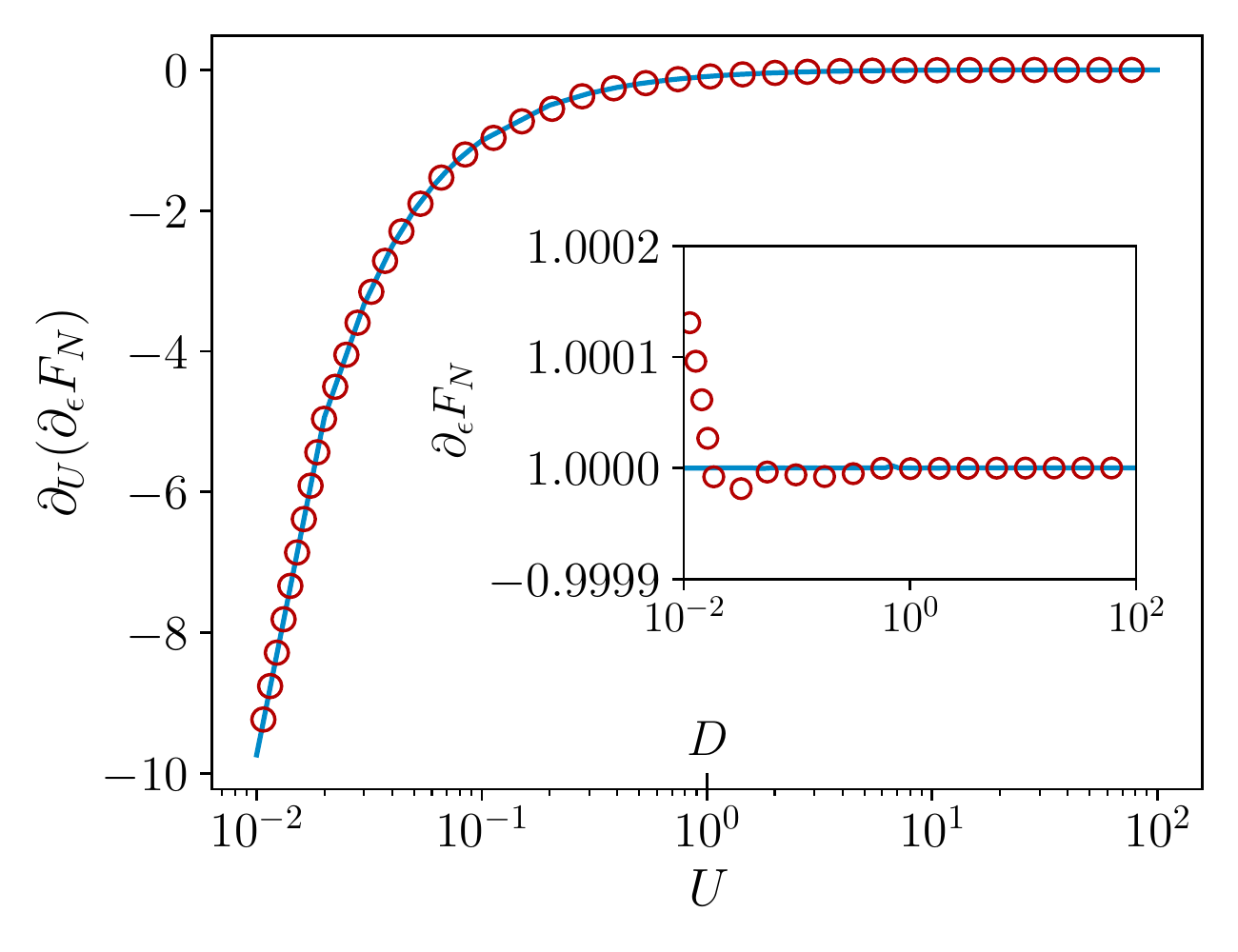}
	\caption{Free energy derivatives of the AIM obtained by $\partial$NRG. Inset shows  $\partial_{\epsilon} F_N \simeq \langle \hat{n}\rangle_T$; main panel shows $\partial_{U}(\partial_{\epsilon} F_N) \simeq \partial_{U}\langle \hat{n}\rangle_T$, comparing AD (blue lines) with FD (red points). Same parameters used as for Fig.~\ref{fig:dGS}, but with $N=20$ and $\bar{\beta}=0.9$ yielding an effective temperature $T/D \simeq 3\times 10^{-5}$.}
	\label{fig:dn}
\end{figure}

\section{Numerical Results}
For the following numerical demonstrations, we implemented a basic NRG code in \textit{jax} \cite{jax2018github}, using the included AD routines to obtain derivatives via \ceqn{eq:mainresult}. The code is available open-source at Ref.~\cite{rigo_mitchell_2021}. For simplicity we did not exploit quantum numbers here, and so eigenstate degeneracies were removed by adding  Gaussian noise with a small variance $\sigma\simeq10^{-11}$ (FD derivatives were calculated for comparison without noise, $\sigma=0$). In the following we set the conduction electron half-bandwidth to $D=1$ and use an NRG discretization parameter $\Lambda=3$. Further details on the numerical calculations and the finite difference derivatives can be found in \capp{app:numerics} and \capp{app:FD} respectively.

First we use $\partial$NRG to compute the derivative of the ground-state energy $E_{N;0}$ of the AIM with respect to the interaction $U$. AD results in \cfig{fig:dGS} (blue lines) are compared with FD approximations (points), as a function of $U$ at iteration $N=5$ (upper panel) and $N=40$ (lower panel), normalized by their respective $U \to 0$ derivatives. The green lines show the variation of $E_{N;0}$ itself. The results show the non-trivial effect of renormalization going from $N=5$ to $N=40$ at intermediate $U$, as well as the saturation of the ground state derivatives at both large and small $U$. In this case we see excellent agreement between AD and FD results (although the former are far less computationally expensive to obtain).

\cfig{fig:dn} demonstrates the use of $\partial$NRG to obtain thermodynamic quantities from derivatives of the NRG free energy. The inset shows the impurity occupation $\langle \hat{n}\rangle \simeq \partial_{\epsilon} F_N$ at a temperature $T/D\sim 10^{-5}$ (corresponding to $N=20$ and $\bar{\beta}=0.9$) for the same systems as in \cfig{fig:dGS}. Since $\epsilon=-U/2$, the AIM possesses an exact particle-hole symmetry and hence is at half-filling, $\langle \hat{n}\rangle =1$. The AD results (blue line) satisfy this exact result precisely, while the FD results (points) show some numerical error. More interestingly, the $\partial$NRG framework allows to obtain higher derivatives with equal ease  (\ceqn{eq:2ndorder} is used instead of \ceqn{eq:deiv}), as shown in the main panel of \cfig{fig:dn}. Here we calculate the corresponding second derivatives $\partial^2 F_N/\partial U\partial \epsilon \simeq \partial_U \langle \hat{n}\rangle$, which again show non-trivial behaviour as a function of interaction strength $U$. The FD approximations agree well, but are much more costly to obtain \cite{grossmann2007numerical}, requiring for every point five executions of the entire NRG code per second derivative, and an expensive convergence test. $\partial$NRG requires only a single function evaluation (see \capp{app:wegenert}). 

\begin{figure}[t]
	\includegraphics[width=\columnwidth]{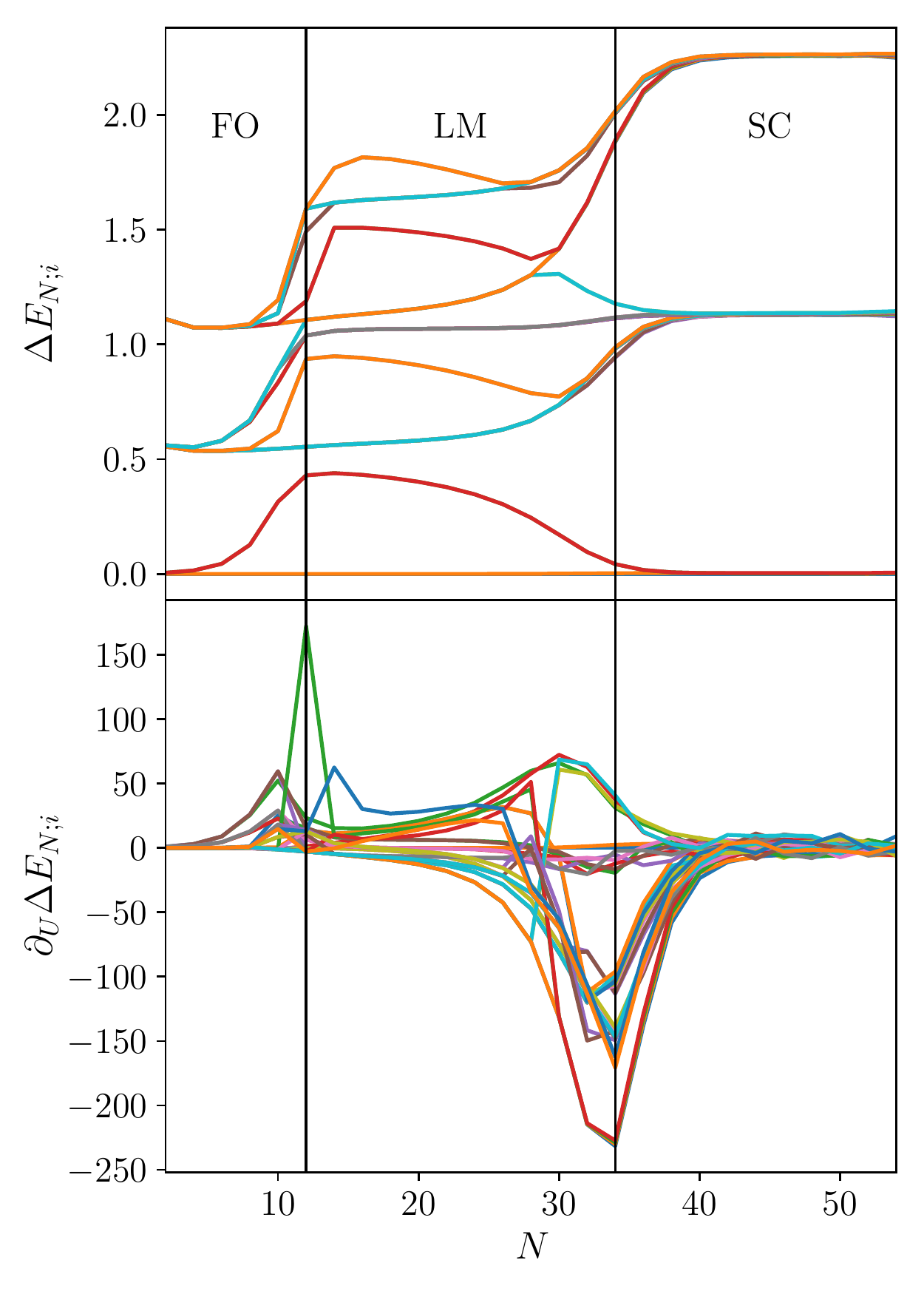}
	\caption{Top panel: NRG energy level flow diagram for the AIM. The lowest 32 rescaled excitation energies $\Delta E_{N;i}$ above the ground state at iteration $N$ are plotted as a function of $N$ (for even $N$ only). Plotted for model parameters $U=0.01$, $\epsilon=-U/2$, and $J\equiv 8V^2/U=0.2$, with the free orbital (FO), local moment (LM) and strong coupling (SC) regimes marked. Bottom panel: the corresponding derivatives with respect to $U$ obtained by $\partial$NRG.}
	\label{fig:dflow}
\end{figure}

As a final application of $\partial$NRG for the AIM, we turn to the RG energy level flow diagram shown in the top panel of \cfig{fig:dflow}. The excitation spectrum of the effective rescaled $\hat{H}_N$ plotted against iteration number $N$ shows the well-known flow between fixed points \cite{wilson1975renormalization,bulla2008numerical,hewson1997kondo}. In the example shown, the crossover scales are well-separated, such that we see distinct level structures associated with the free orbital (FO), local moment (LM), and strong coupling (SC) fixed points as marked on the diagram. At a fixed point, the level structure does not change with $N$. Indeed, the RG structure of the problem and resulting universality implies that the fixed point level structure is the same, independent of the microscopic model parameters $U$ ($\equiv -2\epsilon$) and $V$ -- only the crossover scales between fixed points are affected.

This is demonstrated in the bottom panel of \cfig{fig:dflow}, where we plot the derivatives with respect to $U$ of the NRG energy levels $E_{N;i}$, again as a function of iteration number $N$. As expected, the derivatives vanish at the fixed points (the level structure does not depend on $U$ at the fixed points); but there is a strong dependence along the crossovers, since the crossover scales depend on $U$.


\section{Outlook: dynamics\\and the vertex}
The above numerical results for the AIM are provided as a demonstration proof-of-principle. Future useful applications exploiting the full power of $\partial$NRG may be found for more complex models, situations involving higher-order derivatives, optimization techniques requiring exact gradients, or in cases where partial derivatives of NRG outputs with respect to inputs are difficult to obtain by standard FD means. An example of the latter is the derivative of frequency-dependent dynamical quantities, such as impurity Green's functions or the conduction electron scattering T-matrix, with respect to bare model parameters.

In the context of dynamical mean field theory (DMFT) for correlated materials \cite{georges1996dynamical,kotliar2006electronic,sheridan2021data}, model machine learning techniques \cite{rigo2020machine} could be applied to optimize simpler effective models with respect to more complicated microscopic ones, by comparing their Green's functions. Given the non-trivial dependence of such dynamical quantities at different energies on model parameters, the exact gradient within the AD framework becomes an essential ingredient for gradient descent optimization. 

Refs.~\cite{kugler2021multipoint,lee2021computing} recently uncovered the analytic structure of the full local vertex and proposed a scheme to compute it within NRG. The vertex is an important object, entering for example in extensions of DMFT beyond the local limit \cite{rohringer2018diagrammatic}. It is possible that some partial information on the vertex could also be obtained by $\partial$NRG. This is inspired by the result in Ref.~\cite{van2020bethe} for the AIM which, within the Matsubara formalism, connects the vertex at zero bosonic frequency $F^{\rm loc}_{\nu\nu'}(\omega=0)$ to the functional derivative of the interaction self-energy $\Sigma_{\nu}$ with respect to the hybridization $\Delta_{\nu'}$, viz:
\begin{equation}
\label{eq:Floc}
\frac{\delta\Sigma_{\nu}}{\delta \Delta_{\nu'}} = T [G_{\nu'}]^2  F^{\textrm{loc}}_{\nu \nu'}(\omega=0) \;,
\end{equation}
where $G_{\nu'}$ is the single-particle impurity Matsubara Green's function. For a precise definition and discussion of $F^{\rm loc}_{\nu\nu'}(\omega=0)$, see Ref.~\cite{van2020bethe,kugler2021multipoint}. Since $G_{\nu'}$ and $\Sigma_{\nu}$ can be calculated within standard NRG \cite{osolin2013pade}, we argue that such an object is obtainable within $\partial$NRG. We speculate that a Keldysh version of Eq.~\ref{eq:Floc} may similarly provide access to certain information about the real-frequency vertex at finite temperatures.

In order to calculate such derivatives of dynamical quantities, further code development is required, since the full-density-matrix NRG method would need to be implemented using AD libraries such as \textit{jax} \cite{jax2018github} and integrated within the $\partial$NRG scheme described in this paper. We leave this for future work.


\section{Conclusion}
NRG is the gold-standard method of choice for solving generalized quantum impurity models \cite{wilson1975renormalization,bulla2008numerical}. In this work, we introduce a new variant of the standard algorithm -- $\partial$NRG -- which makes use of the differential programming paradigm to automatically and efficiently obtain derivatives of NRG outputs with respect to input model parameters. 

We make use of the AD \textit{jax} library \cite{jax2018github}, together with a bespoke routine based on \ceqn{eq:mainresult} which allows the derivatives of Hamiltonian eigenvalues and eigenvectors to be obtained at any iteration in an accurate and inexpensive way. Our fully commented code is available open-source \cite{rigo_mitchell_2021}.

We demonstrated the use of $\partial$NRG by application to the Anderson impurity model, for which we obtained the derivative of NRG energy levels with respect to model parameters, calculated thermodynamic quantities from derivatives of the NRG free energy, and  susceptibilities from derivatives of Hamiltonian source terms. $\partial$NRG may be useful for machine learning applications involving NRG \cite{rigo2020machine} for which gradient descent optimization requires derivatives of a loss function; for adaptive broadening schemes \cite{lee2016adaptive}; or for optimal control protocols \cite{vargas2021fully}. Richer physical information may be obtained from derivatives of dynamical quantities. This also opens the door to automatic differentiable DMFT, with $\partial$NRG as the impurity solver. Finally, we note that the $\partial$NRG methodology is compatible with interleaved-NRG (iNRG) \cite{mitchell2014generalized,stadler2016interleaved} and generalizations utilising full symmetries \cite{weichselbaum2012non}. This is left for future work.


\begin{acknowledgments}
	\emph{Acknowledgments.--} 
	The authors thank Lei Wang, Luuk Coopmans, Jan von Delft, Seung-Sup Lee, and Emma Minarelli for the helpful discussions. We acknowledge funding from the Irish Research Council Laureate Awards 2017/2018 through grant IRCLA/2017/169. 
\end{acknowledgments}


\appendix


\begin{figure}
	\centering
	\includegraphics[width=\columnwidth]{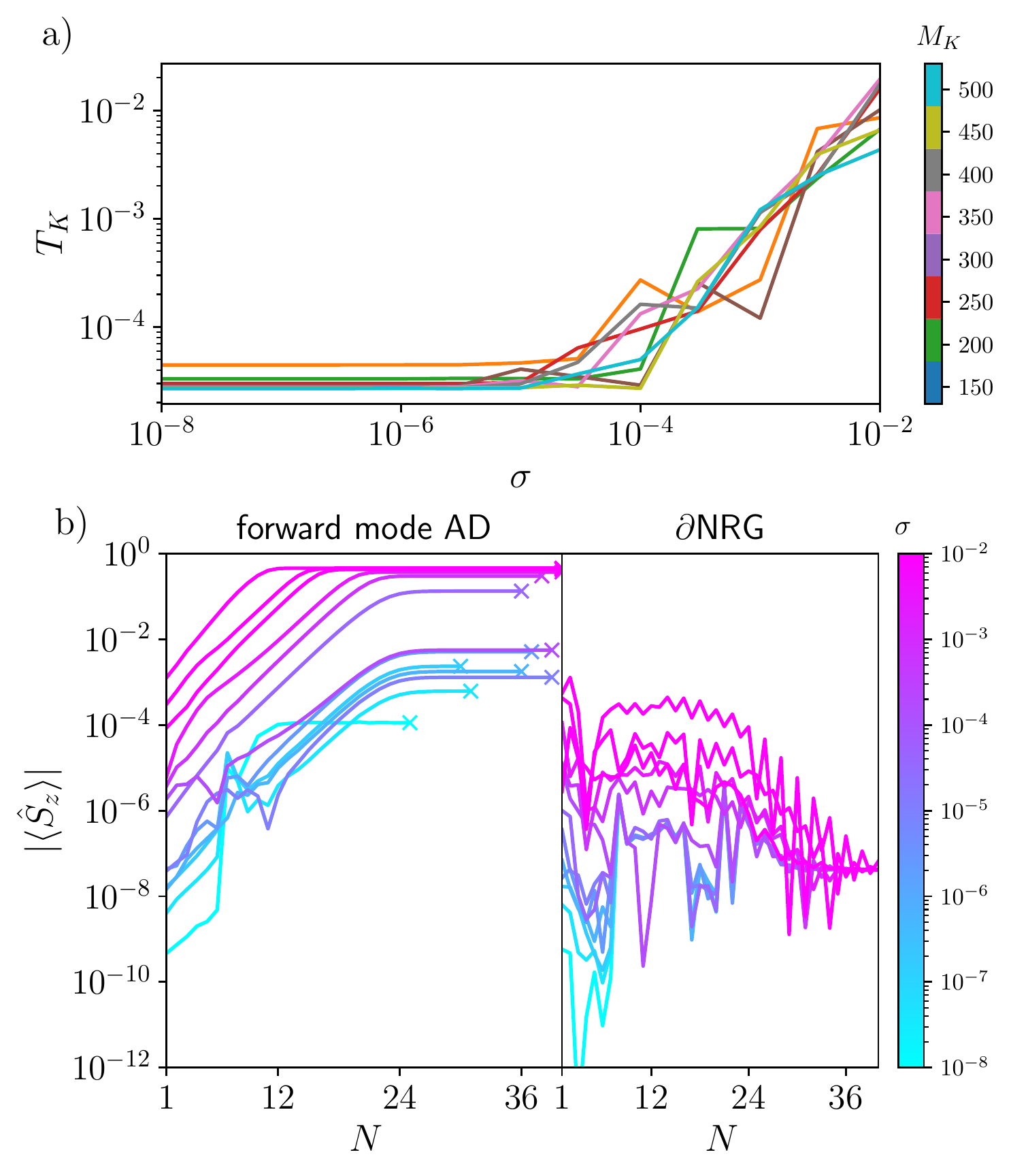}
	\caption{(a) Dependence of $T_K$ on the noise variance $\sigma$ at different $M_K$. The number of kept states ranges from $150$ to $500$ in steps of $50$. The AIM model parameters used are $\epsilon=-0.15,~U=0.3,~V=0.1~B=0$. (b) Zero-field impurity magnetization $\langle \hat{S}_z\rangle$ computed via derivatives of the free energy using $\partial$NRG and forward mode AD, as a function of Wilson chain length $N$ (for $M_K = 500$). The noise variance is $\sigma=1\times10^{-8},3\times10^{-8},1\times10^{-7},3\times10^{-7}...,1\times10^{-2}$. Same model parameters as in (a).}
	\label{fig:acc}
\end{figure}
\twocolumngrid

\section{Numerical benchmarking: accuracy and speedup}
\label{app:benchmark}

In this Appendix we demonstrate that: (i) NRG and  $\partial$NRG are not affected by the addition of Gaussian noise with a small variance; and that (ii) the real-world performance of $\partial$NRG exceeds that of basic AD and FD in terms of both accuracy and speed. 

A degeneracy-free NRG Hamiltonian is required for application of \ceqn{eq:Aderivs}. However, physical systems often do have eigenvalue degeneracies, arising for example from underlying symmetries. Although one can partially mitigate this problem by exploiting Abelian and non-Abelian quantum numbers, some degeneracies may remain. Here we consider the `worst case' scenario in which no symmetries are exploited. 

First we examine the effect of adding Gaussian noise with mean $\mu = 0$ and variance $\sigma$ to the diagonal elements of $\hat{H}_N$. We use the Kondo temperature $T_K$ as one figure-of-merit for assessing the effect of the noise term. For simplicity we define the Kondo temperature through the impurity entropy via  $S_{\rm imp}(T=T_K) = 0.5$ (suitably between the local moment and strong coupling fixed point values). We compute $T_K$ for $10^{-8}\le\sigma\le10^{-2}$ and different numbers of kept states $100 \le M_K \le 500$ (using fixed $\Lambda = 3$). \cfig{fig:acc}(a) shows clearly that for $\sigma \le 10^{-6}$ the Kondo temperature does not noticeably depend on the noise level (we have confirmed explicitly down to $\sigma=10^{-15}$ that the results are fully converged for each $M_K$). Other physical quantities computed in standard NRG show similar behaviour.

However, the effect of adding noise is more pronounced in derivatives. In \cfig{fig:acc}(b) we compare $\partial_B F_N(\bar{\beta}) \vert_{B = 0} = \langle \hat{S}_z \rangle_{H_N,\bar{\beta}}$ as a function of Wilson chain length $N$, as computed with $\partial$NRG and straight AD for different $\sigma$. Since the impurity magnetization is evaluated at zero field, $SU(2)$ spin symmetry implies the exact result $\langle \hat{S}_z \rangle_{H_N,\bar{\beta}} = 0$. However, this spin symmetry is not enforced, and so we see deviations due to the noise. As \cfig{fig:acc}(b) shows,  $\partial$NRG accurately approximates the exact result even at relatively high $\sigma$. By contrast, AD yields  derivatives that strongly depend on $\sigma$ and have much higher error than $\partial$NRG at a given $\sigma$. Indeed, derivatives are numerically not computable for all $N$ in standard forward mode AD for $\sigma \lesssim 10^{-6}$; at larger noise levels, AD derivatives are computable but the accuracy can become poor, especially at later iterations due to the propagation and accumulation of errors through the derivative chain rule.

Therefore, although NRG for the AIM is insensitive to noise for $\sigma \le 10^{-6}$, the AD approach is not stable at these noise levels. At higher noise levels, AD is stabilized, but the accuracy suffers. On the other hand, $\partial$NRG avoids such problems: a lower noise level can be used since each derivative calculation is independent, and highly accurate results can be obtained.

\begin{figure}
	\centering
	\includegraphics[width=\columnwidth]{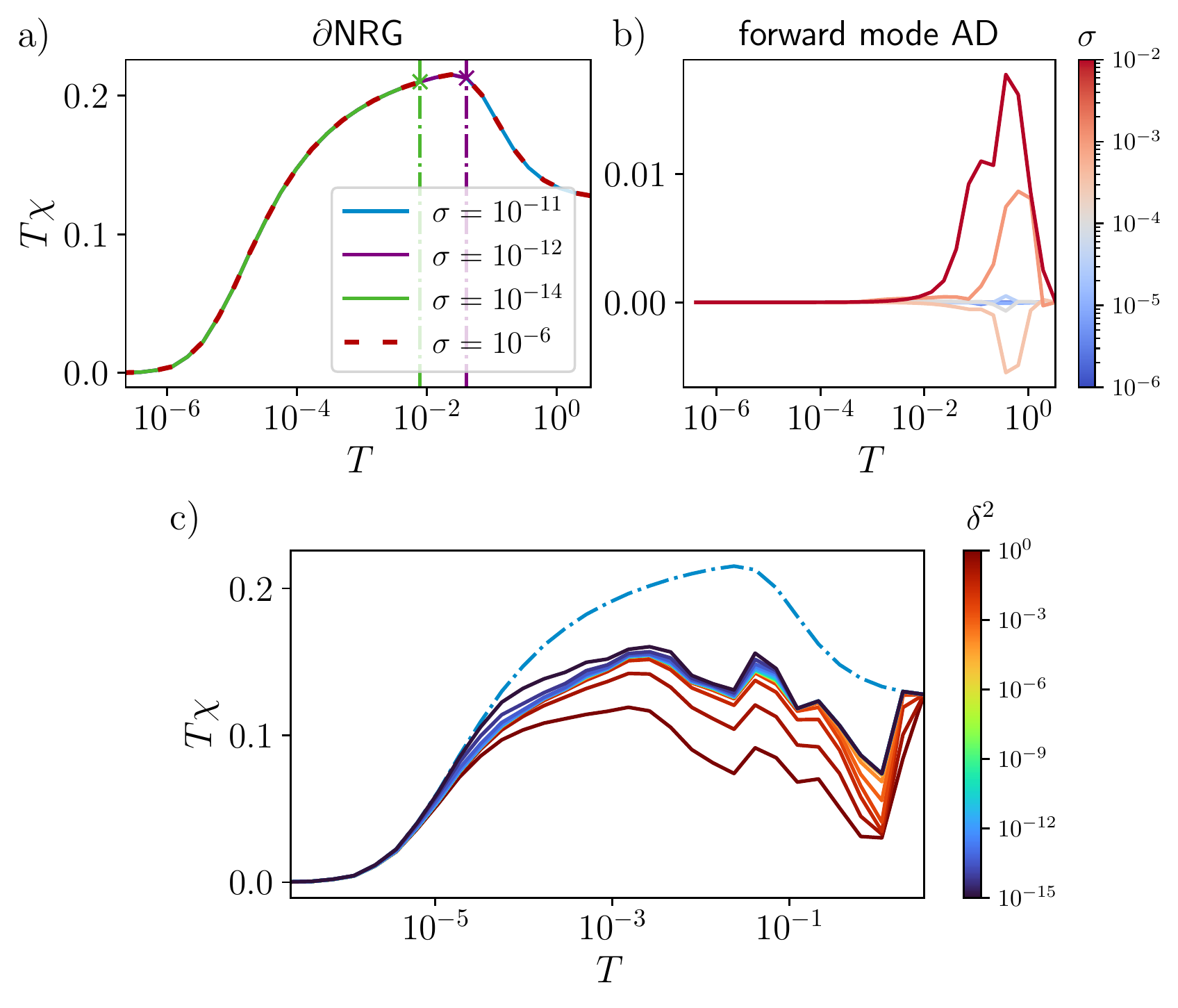}
	\caption{(a) Magnetic susceptibility $\chi$  calculated with $\partial$NRG for different noise variances $\sigma$. (b) $\chi$ computed using forward mode AD and $10^{-8} \le \sigma \le 10^{-2}$. (c) $\chi$ computed with $\partial$NRG using the pole broadening approach, with $10^{-15}\le \delta^2 \le 10^{0}$. All calculations are performed with $M_K =500$ and with the same AIM model parameters as in Fig.~\ref{fig:acc}.}
	\label{fig:acc_chi}
\end{figure}

Within $\partial$NRG, the magnetization calculation via the analog of \ceqn{eq:FN_eps} shown in \cfig{fig:acc}(b) does not involve eigenvector derivatives, and is therefore particularly stable. By contrast, the AD calculation of any derivative is built up recursively via \ceqn{eq:forward} and therefore does involve the computation of eigenvector derivatives at each and every step. This contributes to the relative performance gain in $\partial$NRG.

In \cfig{fig:acc_chi} we examine the magnetic susceptibility $\chi(T)$. The calculation of this quantity, which is obtained automatically in $\partial$NRG via the second derivative of the NRG free energy, is formally equivalent to the analytic expression, \ceqn{eq:chi} (and \textit{does} involve eigenvector derivatives). Gaussian noise of width $\sigma$ is added to the diagonal of $\hat{H}_N$ at iteration $N$. The calculated $\chi_{H_N,\bar{\beta}}$ at $\bar{\beta}=0.9$ (used hereafter) is then interpreted as the true $\chi(T)$ at inverse temperature $\beta = \Lambda^{(N-1)/2}\bar{\beta}$. The numerical results from $\partial$NRG show that $\chi_{H_N,\bar{\beta}}$ is obtained reliably at all $N$ (and hence all $T$) for $\sigma$ as low as $10^{-11}$. Only at smaller $\sigma$ does the method break down: derivatives are then not computable for earlier iterations/higher temperatures (the correct low-$T$ behavior is however still captured). As demonstrated in Fig.~\ref{fig:acc} and confirmed in \cfig{fig:acc_chi}(a), highly accurate results are obtained for $\sigma< 10^{-6}$. Therefore in $\partial$NRG, we have a wide window $10^{-11}<\sigma<10^{-6}$ over which numerical results are fully converged and stable. 

By contrast, in \cfig{fig:acc_chi}(b) we show results for the same quantity obtained by straight AD. The calculation is not numerically stable for small $\sigma$, but very inaccurate for large $\sigma$ (typical of the breakdown for higher-order derivatives). As such there is no reliable regime for which robust results can be obtained by straight AD.

In \cfig{fig:acc_chi}(c) we examine the feasibility of an alternative approach to the eigenvalue-degeneracy problem that avoids adding 
 Gaussian noise. The method, proposed by Liao \textit{et al} in Ref.~\cite{liao2019differentiable}, consists of reformulating \ceqn{eq:deis},
\begin{equation}
\dd\ket{i} = \sum_{i\neq j}\frac{\bra{j} \dd \hat{A} \ket{i}}{\lambda_i - \lambda_j}\ket{j} \equiv \sum_{i\neq j}\bra{j} \dd \hat{A} \ket{i}f(\lambda_i - \lambda_j)\ket{j} \; , 
\end{equation}
with $f(x) = 1/x$. Divergences induced by eigenvalue degeneracies can be avoided by replacing the function $f(x)$ with the approximate broadened form $f(x) \approx x/(x^2 + \delta^2)$ with $\delta \ll 1$. This broadening approximation is known to distort somewhat the overall results, but has the advantage that divergences are removed without the need to lift degeneracies. \cfig{fig:acc_chi}(c) shows numerical results for $\chi(T)$ obtained in this way, as a function of $T$ for different broadenings $\delta$. Although the method always yields a computable result, the true result (dashed line) is only obtained in the $\delta \to 0$ limit at the lowest temperature scales. Finite $\delta$ at intermediate $T$ appears to yield rather poor results. We conclude that $\partial$NRG is indeed the method of choice in this context, in terms of accuracy.\\

\begin{figure}
	\centering
	\includegraphics[width=\columnwidth]{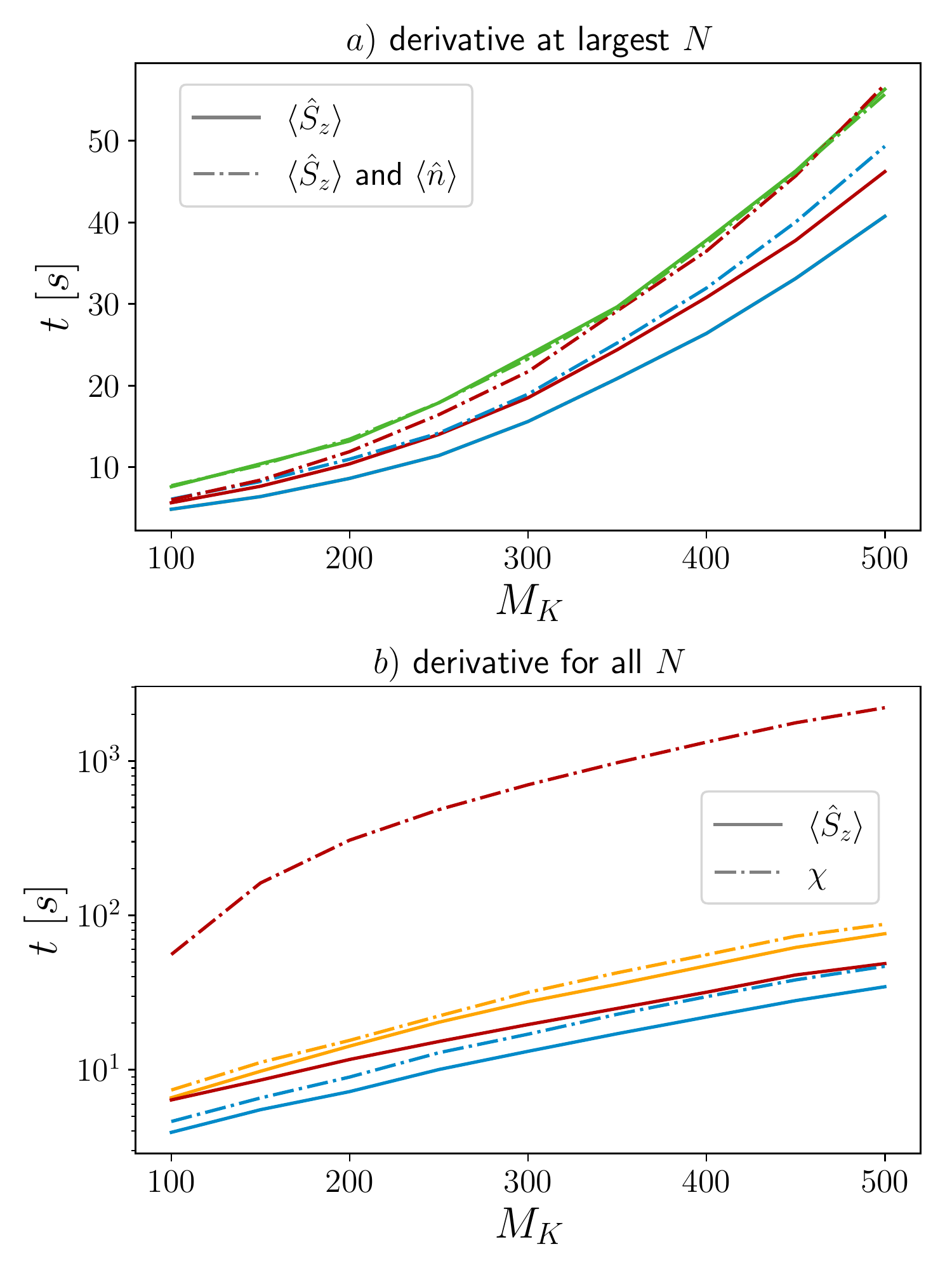}
	\caption{Representative calculation runtime in seconds, vs number of kept states $M_K$, comparing different automatic derivative techniques and computed quantities. In all cases shown, blue lines correspond to $\partial$NRG with single diagonalization, orange lines for $\partial$NRG with double diagonalization, red lines for forward mode AD, and green lines for backward mode AD. 
	(a) Calculation at $N=40$ of $\langle \hat{S}_z \rangle_{H_N,\bar{\beta}}$ (solid lines) compared with the calculation of both $\langle \hat{S}_z \rangle_{H_N,\bar{\beta}}$ and $\langle \hat{n} \rangle_{H_N,\bar{\beta}}$ (dashed lines), obtained via the first derivatives of the NRG free energy. 
	(b) Calculation for all $N\le 40$ of $\langle \hat{S}_z \rangle_{H_N,\bar{\beta}}$ requiring first derivatives (solid lines), and $ \chi_{H_N,\bar{\beta}}$ requiring second derivatives (dashed lines). Model parameters  as in Fig.~\ref{fig:acc}.}
	\label{fig:runtime}
\end{figure}

We now turn to an analysis of the real-world performance of $\partial$NRG in terms of calculation time, comparing with straight AD implemented in \textit{jax}  \cite{jax2018github}. The figure of merit is the runtime measured in seconds $[s]$ of both algorithms run on the same machine (in this case an AMD Threadripper 2950X platform running Python 3.8.10 with the \verb|OMP_NUM_THREAD = 16|
flag; further package specifications can be found in \cite{summerwine}). We do not utilize any quantum numbers here. 

In \cfig{fig:runtime}(a) we compare the runtime for the calculation of (i) $\langle \hat{S}_z \rangle_{H_N,\bar{\beta}}$ (solid lines), and (ii) $\langle \hat{S}_z \rangle_{H_N,\bar{\beta}}$ and $\langle \hat{n} \rangle_{H_N,\bar{\beta}}$ together (dashed lines); using $\partial$NRG (blue),  forward mode AD (red), and backward-mode AD  (green), at the last NRG iteration, $N=40$. We see that for all $M_K$, $\partial$NRG performs best (with the relative speedup becoming more pronounced at larger $M_K$). For straight AD, we see that forward mode beats backward mode for the calculation of a single derivative; but since backward mode AD scales with the number of inputs rather than outputs, it will overtake the  forward mode when many derivatives are calculated. 

In \cfig{fig:runtime}(b) we compare runtimes using $\partial$NRG with a single diagonalization of $\hat{H}_N$ (blue), $\partial$NRG with double diagonalization of $\hat{H}_N$ (orange), and forward mode AD (red). We calculate $\langle \hat{S}_z \rangle_{H_N,\bar{\beta}}$ (solid lines), and $\chi_{H_N,\bar{\beta}}$ (dashed lines) at \emph{all} iterations $N\le 40$. For the orange lines, \emph{two} diagonalizations of $\hat{H}_N$ are performed at each step (with and without noise), which thereby eliminates error propagation from the added noise and provides the most accurate calculation (this may or may not be needed in practice, depending on the situation). 
We see that $\partial$NRG with a single diagonalization per step is the fastest in all cases. For the simpler quantity $\langle \hat{S}_z \rangle_{H_N,\bar{\beta}}$ (which involves only a first-derivative), the $\partial$NRG using two diagonalizations and  forward mode AD have similar runtimes, although AD is slightly faster. However, for $\chi_{H_N,\bar{\beta}}$, both versions of $\partial$NRG are considerably faster. This is because the calculation of $\chi_{H_N,\bar{\beta}}$ requires a second-derivative, which is much more computationally expensive in AD, but almost as cheap in $\partial$NRG. The relative performance gain for $\partial$NRG also increases if several derivatives are computed.

In conclusion, in a typical setting, $\partial$NRG is considerably more efficient than AD (often by orders of magnitude) -- while at the same time being more accurate.


\section{Higher order derivatives}
\label{app:secoder}

\ceqn{eq:Aderivs} allows us to take the derivatives of eigenvalues and eigenvectors of some Hermitian, non-degenerate Hamiltonian $\hat{H}$. Consider now a Hamiltonian that can be linearly decomposed as,
\begin{equation}
\hat{H}(\lbrace\theta\rbrace) = \sum_i \theta_i \hat{h}_i \;,
\end{equation}
where $\hat{h}_i$ is an operator defined over the same Hilbert space as $\hat{H}$, and with the corresponding coupling constant $\theta_i$ a real scalar. In this case, all second order derivatives of the Hamiltonian must vanish,  $\frac{\partial^2 \hat{H}}{\partial \theta_i \partial \theta_j} = 0$. It is then possible to derive a closed-form expression for the second-order derivatives of eigenvalues and eigenvectors with respect to the couplings  $\theta_{\mathbf{1}},\theta_{\mathbf{2}} \in \lbrace\theta\rbrace$, viz:
\begin{subequations}\label{eq:2ndorder}
\begin{align}
\partial_{\theta_{\mathbf{1}}}\partial_{\theta_{\mathbf{2}}} E_i &= 2 \sum_{i \neq j}\frac{ \langle i |  \partial_{\theta_{\mathbf{2}}} \hat{H} \vert j\rangle \langle j |  \partial_{\theta_{\mathbf{1}}} \hat{H} \vert i\rangle}{E_i - E_j} \; ,\\
\partial_{\theta_{\mathbf{1}}}\partial_{\theta_{\mathbf{2}}}|i\rangle &= 
	\sum_{i \neq j}\left[\Pi_{i j}|j\rangle 
	+ \frac{\langle j | \partial_{\theta_{\mathbf{1}}} \hat{H} \vert i\rangle}{\lambda_i - \lambda_j} \sum_{j \neq k} \frac{\langle k | \partial_{\theta_{\mathbf{2}}} \hat{H} \vert j\rangle}{\lambda_j - \lambda_k}\vert k \rangle \right] \;,
\end{align}
\end{subequations}
where,
\begin{widetext}
	\begin{align}
	\Pi_{ij} = -\frac{\langle i|\partial_{\theta_{\mathbf{2}}} \hat{H}|i\rangle - \langle j|\partial_{\theta_{\mathbf{2}}} \hat{H} |j\rangle}{(\lambda_i - \lambda_j)^2}\langle j|\partial_{\theta_{\mathbf{1}}} \hat{H}|i\rangle
	+\sum_{j\neq k}\frac{\langle j | \partial_{\theta_{\mathbf{2}}} \hat{H}| k\rangle}{\lambda_j - \lambda_k}\langle k|\partial_{\theta_{\mathbf{1}}} \hat{H}|i\rangle
	+\sum_{k \neq i}\frac{\langle k |\partial_{\theta_{\mathbf{2}}} \hat{H}|i\rangle}{\lambda_i - \lambda_k}\langle j |\partial_{\theta_{\mathbf{1}}} \hat{H}|k\rangle \; .
	\end{align}
\end{widetext}
With these formulae one can compute physical observables depending on second-derivatives, such as susceptibilities, using $\partial$NRG.


\section{Comparing differentiation methods}
\label{app:wegenert}

\begin{table*}
	\begin{center}
		\begin{tabular}{| p{9cm} | p{9cm} |}
			\hline
			\textbf{Forward Primal Trace} & \textbf{Forward Tangent Trace}  \\
			\hline 
			$v_{-3} = \Lambda$ & $\dd v_{-3} = 0$  \\  
			$v_{-2} = \lbrace t_n \rbrace$ & $\dd v_{-2} = 0$ \\
			$v_{-1} = \eta$ & $\dd v_{-1} = 0$ \\
			$v_{0} = H_{\rm imp}$ & $\dd v_{0} = \dd H_{\rm imp}$ \\
			$v_{1} = \textrm{EIGENVALUES}(v_{0})$ & $ [\dd v_{1}]_i = [ v_{2}]^\dagger_i \cdot \dd v_{0} \cdot [ v_{2}]_i$ \\
			$v_{2} = \textrm{EIGENVECTORS}(v_{0})$ & $ [\dd v_{2}]_i = \sum_{j \neq i}\frac{[ v_{2}]^\dagger_j \cdot \dd v_{0} \cdot [ v_{ 2}]_i }{[v_{1}]_i - [v_{1}]_j}[ v_{2}]_j$ \\
			$v_3 = v_{-1}^T$ & $\dd v_3 = 0$ \\
			FOR $n$ = 0 to $N$ & FOR $n$ = 0 to $N$ \\
			$\qquad v_{7 n + 4} = v_{7 n}$ & $\qquad \dd v_{7n + 4} =  \dd v_{7 n}$ \\
			$\qquad [v_{7n + 5}]_{i i s s} = v_{-3}^{1/2}[v_{7 n + 1}]_{l}$ & $\qquad [\dd v_{7n + 5}]_{i i s s} = v_{-3}^{1/2}[\dd v_{7 n + 2}]_{l}$ \\
			$\qquad [v_{7n + 6}]_{i i' s s'} = (-1)^{i'}[v_{-2}]_n\sum_{\sigma}[v_{7n + 3}]_{\sigma i i'}[v_{-1}]_{\sigma s s'}$ &
			$\qquad [\dd v_{7n + 6}]_{i i' s s'} = (-1)^{i'}[v_{-2}]_n\sum_{\sigma}[\dd v_{7n + 3}]_{\sigma i i'}[v_{-1}]_{\sigma s s'} $ \\
			$\qquad [v_{7n + 7}]_{i i' s s'} = [v_{7n + 5}]_{i i' s s'} + [v_{7n + 6}]_{i i' s s'} + [v_{7n + 6}]^\dagger_{i i' s s'}$ & 
			$\qquad [\dd v_{7 n + 8}]_i = [ v_{7 n + 9}]^\dagger_i \cdot \dd v_{7n} \cdot [ v_{7 n + 9}]_i$  \\
			$\qquad v_{7 n + 8} = \textrm{EIGENVALUES}(v_{7 n + 7 })$ & $\qquad [\dd v_{7 n + 8}]_i = [ v_{7 n + 9}]^\dagger_i \cdot \dd v_{7n} \cdot [ v_{7 n + 9}]_i$ \\
			$\qquad v_{7 n + 9} = \textrm{EIGENVECTORS}(v_{7 n + 7 })$ &
			$\qquad [\dd v_{7 n + 9}]_i = \sum_{j \neq i}\frac{[ v_{7 n + 9}]^\dagger_j \cdot \dd v_{7n} \cdot [ v_{7 n + 9}]_i }{[v_{7 n + 8}]_i - [v_{7 n + 8}]_j}[ v_{7 n + 9}]_j$ \\
			$\qquad [v_{7n + 10}]_{\sigma i i'} =\sum_{s s'}\sum_{j j'} [v_{7 n + 9}]^\dagger_{i;s j}[v_{7 n + 9}]_{i;s' j'} [v_{-1}]_{\sigma s s'} $ &
			$\qquad [\dd v_{7n + 10}]_{\sigma i i'} = \sum_{s s'}\sum_{j j'} \left( [\dd v_{7 n + 9}]^\dagger_{i;s j}[v_{7 n + 9}]_{i;s' j'}  \right.$ \newline 
			$\qquad~~~~~~~  + \left.[ v_{7 n + 9}]^\dagger_{i;s j}[\dd v_{7 n + 9}]_{i;s' j'} \right)[v_{-1}]_{\sigma s s'}$ \\
			$y_0 = v_{7N+7}$ & $\dd y_0 = \dd v_{7N+7}$ \\		
			\hline
		\end{tabular}
		\caption{Wengert list for NRG with  forward mode AD. Left column: the primal trace prescribes all steps to be performed to execute an NRG calculation with $N$ iterations. Right column: the tangent trace prescribes all steps to compute the derivative the NRG Hamiltonian $\hat{H}_N$.}
		\label{tab:forwardmode}
	\end{center}
\end{table*}

\begin{table*}
	\begin{center}
		\begin{tabular}{| p{10cm} | p{4.5cm} |}
			\hline
			\textbf{Forward Primal Trace} & \textbf{Forward Tangent Trace}  \\
			\hline
			$v_{-4} =  \dd H_{\rm imp}$ & $dv_{-4} =  \dd H_{\rm imp}$ \\ 
			$v_{-3} = \Lambda$ & $\dd v_{-3} = 0$ \\ 
			$v_{-2} = \lbrace t_n \rbrace$ & $\dd v_{-2} = 0$ \\ 
			$v_{-1} = \eta$ & $\dd v_{-1} = 0$ \\ 
			$v_{0} = H_{\rm imp}$ & $dv_{-4} =  0$ \\
			$v_{1} = \textrm{EIGENVALUES}(v_{0})$ & \\
			$v_{2} = \textrm{EIGENVECTORS}(v_{0})$ & \\
			$v_3 = v_{-1}^T$ & \\
			$[v_4]_{ii'} = \sum_{s s'}\sum_{j j'} [v_{2}]^\dagger_{i;s j}[v_{2}]_{i;s' j'} [v_{-4}]_{jj'} \delta_{s s'}$ & \\
			FOR $n$ = 0 to $N$ & \\
			$\qquad v_{8 n + 5} = v_{8 n}$ & \\
			$\qquad [v_{8n + 6}]_{i i s s} = v_{-3}^{1/2}[v_{8 n + 1}]_{l}$ & \\
			$\qquad [v_{8n + 7}]_{i i' s s'} = (-1)^{i'}[v_{-2}]_n\sum_{\sigma}[v_{8n + 3}]_{\sigma i i'}[v_{-1}]_{\sigma s s'}$ & \\
			$\qquad [v_{8n + 8}]_{i i' s s'} = [v_{8n + 6}]_{i i' s s'} + [v_{8n + 7}]_{i i' s s'} + [v_{8n + 7}]^\dagger_{i i' s s'}$ & \\
			$\qquad v_{8 n + 9} = \textrm{EIGENVALUES}(v_{8 n + 8 })$ & \\
			$\qquad v_{8 n + 10} = \textrm{EIGENVECTORS}(v_{8 n + 8 })$ & \\
			$\qquad [v_{8n + 11}]_{\sigma i i'} =\sum_{s s'}\sum_{j j'} [v_{8 n + 10}]^\dagger_{i;s j}[v_{8 n + 10}]_{i;s' j'} [v_{-1}]_{\sigma s s'} $ & \\
			$\qquad [v_{8n+12}]_{ii'} = \sum_{s s'}\sum_{j j'} [v_{8n+10}]^\dagger_{i;s j}[v_{8n+10}]_{i;s' j'} [v_{8n+4}]_{jj'} \delta_{s s'}$ & \\
			$y_0 = v_{8N+8}$ & $\dd y_0 = v_{8N+12}$ \\
			\hline
		\end{tabular}
		\caption{Wengert list for $\partial$NRG. Left column: the forward primal trace for $\partial$NRG outputs the NRG Hamiltonain $\hat{H}_N$ and its derivative $d\hat{H}_N$. Since derivatives are computed via Eq.~(\ref{eq:mainresult}) in the main algorithm, there is no corresponding forward tangent trace in the Right column.}
		\label{tab:dNRG}
	\end{center}
\end{table*}

The basic NRG algorithm is described in the main text, with the key steps contained in \ceqn{eq:wc}-(\ref{eq:newH}). For more details, see Ref.~\cite{bulla2008numerical}. Assuming that degeneracies in the NRG Hamiltonian $\hat{H}_N$ are lifted, we can use \ceqn{eq:Aderivs} to compute the derivatives of eigenvectors and eigenvalues of $\hat{H}_N$. The Wengert list \cite{wengert1964simple,baydin2018automatic} with the forward primal trace (function evaluation) and the forward tangent trace (function differentiation) can then be compiled, as shown in \ctab{tab:forwardmode}. Similarly we compile the forward primal trace and forward tangent trace for the $\partial$NRG algorithm, see \ctab{tab:dNRG}. For $\partial$NRG, we have one additional step in the primal trace for the main algorithm (corresponding to \ceqn{eq:Ptrafo}), but a trivial tangent trace.

We can now compare the efficiency of the two approaches. For forward mode AD (fAD) we have $\ops(\textrm{fAD}) = 7 N$, and to compute forward primal and tangent traces, $2\times\ops(\textrm{fAD})$ operations are required. For $\partial$NRG by contrast, $\ops(\textrm{$\partial$NRG}) = 8 N$, but no other step is required to compute the derivative of $\hat{H}_N$. However, this does not mean that $\partial$NRG is twice as fast as fAD because the bottleneck eigensolver routines \cite{demmel2008performance} appear only in the forward primal trace. Nonetheless $\partial$NRG still provides a considerable performance advantage, as established by explicit benchmarks in \capp{app:benchmark}

\begin{figure}[t!]
	\centering
	\includegraphics[width=\columnwidth]{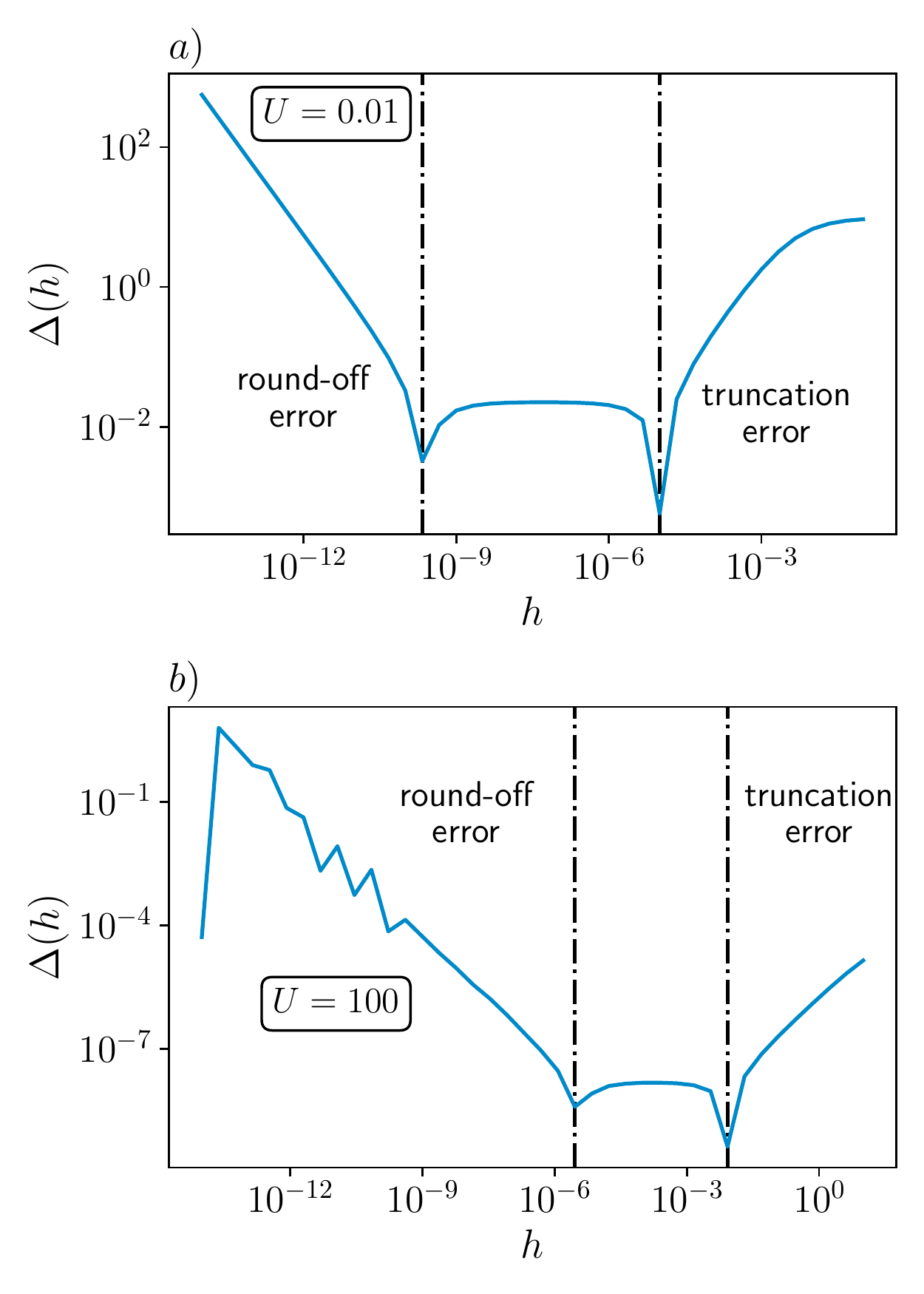}
	\caption{FD derivative errors $\Delta(h)$ vs FD value $h$ for the calculation of $\partial_U\langle \hat{n} \rangle_{H_N,\bar{\beta}}$ at $N=40$. Shown for AIM model parameters $\epsilon = - U/2$, $V=0.1$ and $B=0$ with (a) $U=0.01$, and (b) $U=100$. In all cases $M_K = 500$.}
	\label{fig:FD}
\end{figure}


\section{Details of numerics - NRG calculations}
\label{app:numerics}

We used standard thermodynamic NRG \cite{wilson1975renormalization} to calculate the free energy, impurity occupation and eigenvalues of $\hat{H}_N$. An NRG discretization parameter $\Lambda=3$ was used, and  $M_K=200$ states were retained at each step for Figs.~(1) and (3), while  $M_K=600$ states were retained for Fig.~(2). No quantum numbers were used in this demonstration calculation. 


\section{Details of finite difference calculations}
\label{app:FD}

We calculate the finite difference (FD) derivative of a function $f$ via,
\begin{equation}
\label{eq:FD1app}
D_h[f](x)=\frac{f(x+h) -f(x)}{h}  \; ,
\end{equation}
where $h$ is the FD value. The FD derivative is connected to the definition of the derivative by taking the limit, $\frac{\partial f}{\partial x} = \lim_{h \to 0} D_h[f](x)$.

In practice we compute FD derivatives for a set of finite $\lbrace h_i \rbrace$, and check for convergence. However, care must be taken due to the trade-off between \textit{round-off} and \textit{truncation} errors \cite{baydin2018automatic}. The truncation error arises due to the finite $h > 0$, which is required for the numerical evaluation of $D_h[f](x)$, and which would diminish in the formal limit $h \rightarrow 0$. However for very small $h$, the difference in the function evaluations for $f(x)$ and $f(x+h)$ cannot be distinguished numerically due to inevitable round-off errors in any finite precision arithmetic. This leads to increasing errors as $h \rightarrow 0$. 

We quantify  the FD error as,
\begin{equation}\label{eq:fderror}
	\Delta(h) = \left\vert D_h[f](x) -\frac{\partial f}{\partial x} \right\vert \;.
\end{equation}
This quantity is plotted in \cfig{fig:FD} for the derivative of the AIM impurity occupation with respect to the impurity interaction, $\partial_U\langle \hat{n} \rangle_{H_N,\bar{\beta}}$. We have used the $\partial$NRG derivative as the true value of $\frac{\partial f}{\partial x}$ in \ceqn{eq:fderror}. 
 \cfig{fig:FD} shows rather typical behavior, with truncation errors dominating at large $h$ and round-off errors dominating at small $h$, with a stability plateau in between where the derivative should be evaluated. However, comparison of the upper and lower panels (which correspond to model parameters  $U=10^{-2}$ and $U=10^2$ respectively), demonstrates an important limitation of FD differentiation: the optimal $h$ is not fixed but depends on input model parameters. Reliable results therefore require such a convergence analysis for each new set of model parameters and for each new derivative. This becomes computationally costly. The situation becomes significantly worse when considering higher-order derivatives.

\vfill


\bibliography{refs}

\end{document}